\documentstyle[10pt,epsf,epsfig,dp_delphititle,oldlfont]{dp_delphi}
\makeindex
\def\DpPaperGroup{EP}
\def\DpPaperRef{2001-030}
\def\DpDate{09 April 2001}
\def\DpAuthors{DELPHI Collaboration}
\def\DpTitle{A Measurement of the Tau Topological Branching Ratios}
\def\DpSubmit{(Accepted by Eur.Phys.J. C)}
\def\DpComment{}
\def\DpEMail{}

\begin{document}
\makeatletter
\newcount\@tempcntc
\def\@citex[#1]#2{\if@filesw\immediate\write\@auxout{\string\citation{#2}}\fi
  \@tempcnta\z@\@tempcntb\m@ne\def\@citea{}\@cite{\@for\@citeb:=#2\do
    {\@ifundefined
       {b@\@citeb}{\@citeo\@tempcntb\m@ne\@citea\def\@citea{,}{\bf ?}\@warning
       {Citation `\@citeb' on page \thepage \space undefined}}%
    {\setbox\z@\hbox{\global\@tempcntc0\csname b@\@citeb\endcsname\relax}%
     \ifnum\@tempcntc=\z@ \@citeo\@tempcntb\m@ne
       \@citea\def\@citea{,}\hbox{\csname b@\@citeb\endcsname}%
     \else
      \advance\@tempcntb\@ne
      \ifnum\@tempcntb=\@tempcntc
      \else\advance\@tempcntb\m@ne\@citeo
      \@tempcnta\@tempcntc\@tempcntb\@tempcntc\fi\fi}}\@citeo}{#1}}
\def\@citeo{\ifnum\@tempcnta>\@tempcntb\else\@citea\def\@citea{,}%
  \ifnum\@tempcnta=\@tempcntb\the\@tempcnta\else
   {\advance\@tempcnta\@ne\ifnum\@tempcnta=\@tempcntb \else \def\@citea{--}\fi
    \advance\@tempcnta\m@ne\the\@tempcnta\@citea\the\@tempcntb}\fi\fi}
 
\makeatother
\begin{titlepage}
\pagenumbering{roman}
\CERNpreprint{\DpPaperGroup}{\DpPaperRef} 
\date{{\small\DpDate}} 
\title{\DpTitle} 
\address{\DpAuthors} 
\begin{shortabs} 
\noindent
    Using data collected  in the DELPHI  detector at  LEP-1,
    measurements of the inclusive $\tau$ branching ratios
    for decay modes containing one, three, or five charged particles have been performed,
    giving the following results:\\
\[\begin{array}{llcr}
B_1 & \equiv B(\tau^- \to {\mathrm (particle)}^-  
      \!\geq\!0\pi^0 \!\geq\!0K^0\nu_\tau(\bar{\nu})) & =&(85.316\pm0.093\pm0.049)\%;\\
B_3 & \equiv B(\tau^- \to 2h^-h^+  \!\geq\!0\pi^0 \!\geq\!0K^0\nu_\tau) & =&(14.569\pm0.093\pm0.048 )\%;\\
B_5 & \equiv B(\tau^- \to 3h^-2h^+ \!\geq\!0\pi^0 \!\geq\!0K^0\nu_\tau) & =&(0.115\pm0.013\pm0.006)\%,
\end{array}\]
    where $h$ is either a charged $\pi$ or $K$ meson.
    The first quoted uncertainties are statistical and the second systematic.

\end{shortabs}
\vfill
\begin{center}
\DpSubmit \ \\ 
\DpComment \ \\
\DpEMail \ \\
\end{center}
\vfill
\clearpage
\headsep 10.0pt
\addtolength{\textheight}{10mm}
\addtolength{\footskip}{-5mm}
\begingroup
%
\newcommand{\DpName}[2]{\hbox{#1$^{\ref{#2}}$},\hfill}
\newcommand{\DpNameTwo}[3]{\hbox{#1$^{\ref{#2},\ref{#3}}$},\hfill}
\newcommand{\DpNameThree}[4]{\hbox{#1$^{\ref{#2},\ref{#3},\ref{#4}}$},\hfill}
\newskip\Bigfill \Bigfill = 0pt plus 1000fill
\newcommand{\DpNameLast}[2]{\hbox{#1$^{\ref{#2}}$}\hspace{\Bigfill}}
%
\footnotesize
\noindent
\DpName{P.Abreu}{LIP}
\DpName{W.Adam}{VIENNA}
\DpName{T.Adye}{RAL}
\DpName{P.Adzic}{DEMOKRITOS}
\DpName{Z.Albrecht}{KARLSRUHE}
\DpName{T.Alderweireld}{AIM}
\DpName{G.D.Alekseev}{JINR}
\DpName{R.Alemany}{CERN}
\DpName{T.Allmendinger}{KARLSRUHE}
\DpName{P.P.Allport}{LIVERPOOL}
% ...
\normalsize
\endgroup
\titlefoot{Department of Physics and Astronomy, Iowa State
     University, Ames IA 50011-3160, USA
    \label{AMES}}
\titlefoot{Physics Department, Univ. Instelling Antwerpen,
     Universiteitsplein 1, B-2610 Antwerpen, Belgium \\
     \indent~~and IIHE, ULB-VUB,
     Pleinlaan 2, B-1050 Brussels, Belgium \\
     \indent~~and Facult\'e des Sciences,
     Univ. de l'Etat Mons, Av. Maistriau 19, B-7000 Mons, Belgium
    \label{AIM}}
\titlefoot{Physics Laboratory, University of Athens, Solonos Str.
     104, GR-10680 Athens, Greece
    \label{ATHENS}}
\titlefoot{Department of Physics, University of Bergen,
     All\'egaten 55, NO-5007 Bergen, Norway
    \label{BERGEN}}
\titlefoot{Dipartimento di Fisica, Universit\`a di Bologna and INFN,
     Via Irnerio 46, IT-40126 Bologna, Italy
    \label{BOLOGNA}}
\titlefoot{Centro Brasileiro de Pesquisas F\'{\i}sicas, rua Xavier Sigaud 150,
     BR-22290 Rio de Janeiro, Brazil \\
     \indent~~and Depto. de F\'{\i}sica, Pont. Univ. Cat\'olica,
     C.P. 38071 BR-22453 Rio de Janeiro, Brazil \\
     \indent~~and Inst. de F\'{\i}sica, Univ. Estadual do Rio de Janeiro,
     rua S\~{a}o Francisco Xavier 524, Rio de Janeiro, Brazil
    \label{BRASIL}}
\titlefoot{Comenius University, Faculty of Mathematics and Physics,
     Mlynska Dolina, SK-84215 Bratislava, Slovakia
    \label{BRATISLAVA}}
\titlefoot{Coll\`ege de France, Lab. de Physique Corpusculaire, IN2P3-CNRS,
     FR-75231 Paris Cedex 05, France
    \label{CDF}}
\titlefoot{CERN, CH-1211 Geneva 23, Switzerland
    \label{CERN}}
\titlefoot{Institut de Recherches Subatomiques, IN2P3 - CNRS/ULP - BP20,
     FR-67037 Strasbourg Cedex, France
    \label{CRN}}
\titlefoot{Now at DESY-Zeuthen, Platanenallee 6, D-15735 Zeuthen, Germany
    \label{DESY}}
\titlefoot{Institute of Nuclear Physics, N.C.S.R. Demokritos,
     P.O. Box 60228, GR-15310 Athens, Greece
    \label{DEMOKRITOS}}
\titlefoot{FZU, Inst. of Phys. of the C.A.S. High Energy Physics Division,
     Na Slovance 2, CZ-180 40, Praha 8, Czech Republic
    \label{FZU}}
\titlefoot{{Now at D.P.N.C., University of Geneva, Quai Ernest-Ansermet 24, CH-1211 Geneva 4, Switzerland}
    \label{GENEVA}}
\titlefoot{Dipartimento di Fisica, Universit\`a di Genova and INFN,
     Via Dodecaneso 33, IT-16146 Genova, Italy
    \label{GENOVA}}
\titlefoot{Institut des Sciences Nucl\'eaires, IN2P3-CNRS, Universit\'e
     de Grenoble 1, FR-38026 Grenoble Cedex, France
    \label{GRENOBLE}}
\titlefoot{Helsinki Institute of Physics, HIP,
     P.O. Box 9, FI-00014 Helsinki, Finland
    \label{HELSINKI}}
\titlefoot{Joint Institute for Nuclear Research, Dubna, Head Post
     Office, P.O. Box 79, RU-101 000 Moscow, Russian Federation
    \label{JINR}}
\titlefoot{Institut f\"ur Experimentelle Kernphysik,
     Universit\"at Karlsruhe, Postfach 6980, DE-76128 Karlsruhe,
     Germany
    \label{KARLSRUHE}}
\titlefoot{Institute of Nuclear Physics and University of Mining and Metalurgy,
     Ul. Kawiory 26a, PL-30055 Krakow, Poland
    \label{KRAKOW}}
\titlefoot{Universit\'e de Paris-Sud, Lab. de l'Acc\'el\'erateur
     Lin\'eaire, IN2P3-CNRS, B\^{a}t. 200, FR-91405 Orsay Cedex, France
    \label{LAL}}
\titlefoot{School of Physics and Chemistry, University of Lancaster,
     Lancaster LA1 4YB, UK
    \label{LANCASTER}}
\titlefoot{LIP, IST, FCUL - Av. Elias Garcia, 14-$1^{o}$,
     PT-1000 Lisboa Codex, Portugal
    \label{LIP}}
\titlefoot{Department of Physics, University of Liverpool, P.O.
     Box 147, Liverpool L69 3BX, UK
    \label{LIVERPOOL}}
\titlefoot{LPNHE, IN2P3-CNRS, Univ.~Paris VI et VII, Tour 33 (RdC),
     4 place Jussieu, FR-75252 Paris Cedex 05, France
    \label{LPNHE}}
\titlefoot{Department of Physics, University of Lund,
     S\"olvegatan 14, SE-223 63 Lund, Sweden
    \label{LUND}}
\titlefoot{Universit\'e Claude Bernard de Lyon, IPNL, IN2P3-CNRS,
     FR-69622 Villeurbanne Cedex, France
    \label{LYON}}
\titlefoot{Univ. d'Aix - Marseille II - CPP, IN2P3-CNRS,
     FR-13288 Marseille Cedex 09, France
    \label{MARSEILLE}}
\titlefoot{Dipartimento di Fisica, Universit\`a di Milano and INFN-MILANO,
     Via Celoria 16, IT-20133 Milan, Italy
    \label{MILANO}}
\titlefoot{Dipartimento di Fisica, Univ. di Milano-Bicocca and
     INFN-MILANO, Piazza delle Scienze 2, IT-20126 Milan, Italy
    \label{MILANO2}}
\titlefoot{IPNP of MFF, Charles Univ., Areal MFF,
     V Holesovickach 2, CZ-180 00, Praha 8, Czech Republic
    \label{NC}}
\titlefoot{NIKHEF, Postbus 41882, NL-1009 DB
     Amsterdam, The Netherlands
    \label{NIKHEF}}
\titlefoot{National Technical University, Physics Department,
     Zografou Campus, GR-15773 Athens, Greece
    \label{NTU-ATHENS}}
\titlefoot{Physics Department, University of Oslo, Blindern,
     NO-1000 Oslo 3, Norway
    \label{OSLO}}
\titlefoot{Dpto. Fisica, Univ. Oviedo, Avda. Calvo Sotelo
     s/n, ES-33007 Oviedo, Spain
    \label{OVIEDO}}
\titlefoot{Department of Physics, University of Oxford,
     Keble Road, Oxford OX1 3RH, UK
    \label{OXFORD}}
\titlefoot{Dipartimento di Fisica, Universit\`a di Padova and
     INFN, Via Marzolo 8, IT-35131 Padua, Italy
    \label{PADOVA}}
\titlefoot{Rutherford Appleton Laboratory, Chilton, Didcot
     OX11 OQX, UK
    \label{RAL}}
\titlefoot{Dipartimento di Fisica, Universit\`a di Roma II and
     INFN, Tor Vergata, IT-00173 Rome, Italy
    \label{ROMA2}}
\titlefoot{Dipartimento di Fisica, Universit\`a di Roma III and
     INFN, Via della Vasca Navale 84, IT-00146 Rome, Italy
    \label{ROMA3}}
\titlefoot{DAPNIA/Service de Physique des Particules,
     CEA-Saclay, FR-91191 Gif-sur-Yvette Cedex, France
    \label{SACLAY}}
\titlefoot{Instituto de Fisica de Cantabria (CSIC-UC), Avda.
     los Castros s/n, ES-39006 Santander, Spain
    \label{SANTANDER}}
\titlefoot{Dipartimento di Fisica, Universit\`a degli Studi di Roma
     La Sapienza, Piazzale Aldo Moro 2, IT-00185 Rome, Italy
    \label{SAPIENZA}}
\titlefoot{Inst. for High Energy Physics, Serpukov
     P.O. Box 35, Protvino, (Moscow Region), Russian Federation
    \label{SERPUKHOV}}
\titlefoot{J. Stefan Institute, Jamova 39, SI-1000 Ljubljana, Slovenia
     and Laboratory for Astroparticle Physics,\\
     \indent~~Nova Gorica Polytechnic, Kostanjeviska 16a, SI-5000 Nova Gorica, Slovenia, \\
     \indent~~and Department of Physics, University of Ljubljana,
     SI-1000 Ljubljana, Slovenia
    \label{SLOVENIJA}}
\titlefoot{Fysikum, Stockholm University,
     Box 6730, SE-113 85 Stockholm, Sweden
    \label{STOCKHOLM}}
\titlefoot{Dipartimento di Fisica Sperimentale, Universit\`a di
     Torino and INFN, Via P. Giuria 1, IT-10125 Turin, Italy
    \label{TORINO}}
\titlefoot{Dipartimento di Fisica, Universit\`a di Trieste and
     INFN, Via A. Valerio 2, IT-34127 Trieste, Italy \\
     \indent~~and Istituto di Fisica, Universit\`a di Udine,
     IT-33100 Udine, Italy
    \label{TU}}
\titlefoot{Univ. Federal do Rio de Janeiro, C.P. 68528
     Cidade Univ., Ilha do Fund\~ao
     BR-21945-970 Rio de Janeiro, Brazil
    \label{UFRJ}}
\titlefoot{Department of Radiation Sciences, University of
     Uppsala, P.O. Box 535, SE-751 21 Uppsala, Sweden
    \label{UPPSALA}}
\titlefoot{IFIC, Valencia-CSIC, and D.F.A.M.N., U. de Valencia,
     Avda. Dr. Moliner 50, ES-46100 Burjassot (Valencia), Spain
    \label{VALENCIA}}
\titlefoot{Institut f\"ur Hochenergiephysik, \"Osterr. Akad.
     d. Wissensch., Nikolsdorfergasse 18, AT-1050 Vienna, Austria
    \label{VIENNA}}
\titlefoot{Inst. Nuclear Studies and University of Warsaw, Ul.
     Hoza 69, PL-00681 Warsaw, Poland
    \label{WARSZAWA}}
\titlefoot{Fachbereich Physik, University of Wuppertal, Postfach
     100 127, DE-42097 Wuppertal, Germany
    \label{WUPPERTAL}}
\addtolength{\textheight}{-10mm}
\addtolength{\footskip}{5mm}
\clearpage
\headsep 30.0pt
\end{titlepage}
%
\pagenumbering{arabic} 
\setcounter{footnote}{0} %
\large
\newcommand{\qcth}{\cos\!\theta}
\newcommand{\AVHC}{E_{hlay}}
\newcommand{\GeVA}{\mbox{\rm GeV}}
\newcommand{\GeVB}{\mbox{\rm GeV}/$c$}
\newcommand{\GeVC}{\mbox{\rm GeV}/$c^2$}
\newcommand{\MeVA}{\mbox{\rm MeV}}
\newcommand{\MeVB}{\mbox{\rm MeV}/$c$}  
\newcommand{\MeVC}{\mbox{\rm MeV}/$c^2$}
\newcommand{\pnoOD}{p'}
\newcommand{\pbeam}{p_{beam}}
\newcommand{\ebeam}{E_{beam}}
\newcommand{\ptau}{{\cal P}_{_{\scriptstyle \!\tau}}}
\newcommand{\ptauav}{\langle \ptau \rangle}
\newcommand{\pz}{{\cal P}_{_{\!\!\mbox{\tiny Z}}}}
\newcommand{\Ecm}{E_{cm}}
\newcommand{\Bone}{B_1}
\newcommand{\Bthree}{B_3}
\newcommand{\Bfive}{B_5}
\def\mtau{m_\tau}
\def\mtau2{m^2_{\tau}}
\def\sinw{\sin^2\theta_W}
\def\elec{\rm e}
\def\ee{ {\rm e}^+ {\rm e}^-}
\def\tt{ \tau^+\tau^-}
\def\mm{ \mu^+\mu^-}
\def\ll{l^+l^-}
\def\qq{ {\rm q} \bar{\rm q}}
\def\ff{ {\rm f} \bar{\rm f}}
\def\gg{\gamma\gamma}
\def\eett{\ee \rightarrow \tt}
\def\eeztt{\ee \rightarrow {\rm Z} \rightarrow \tt}
\def\eemm{\ee \rightarrow \mm}
\def\bhab{\ee \rightarrow \ee}
\def\eeha{\ee \rightarrow \qq}
\def\ggee{\ee \rightarrow \ee\ee}
\def\ggmm{\ee \rightarrow \ee\mm}
\def\ggtt{\ee \rightarrow \ee\tt}
\newcommand{\aone}{{\rm a}_1}
\def\a1nu{\aone\nu}
\def\GVA{\gamma_{VA}}
\def\pull{\Pi_{dE/dx}}
\def\pullep{\Pi_{E/p}}
\def\pel{p_{el}}
\def\EMM{E_{3X_0}}
\def\BWr{BW_{\rho}}
\def\fvx{f(\vec{x})}
\def\gvx{g(\vec{x})}
\def\vx{{\bf \vec{x}}}
\def\pp{{\bf \vec{p}}}
\def\ppi{{\bf \vec{p}_i}}
\def\ppj{{\bf \vec{p}_j}}
\def\ppk{{\bf \vec{p}_k}}
\def\pu{{\bf \vec{p}_1}}
\def\ptp{{\bf \vec{p}_{3\pi}}}
\def\pd{{\bf \vec{p}_2}}
\def\pt{{\bf \vec{p}_3}}
\def\ct{\cos{\theta_h}}
\def\cdt{\cos^2{\theta}}
\def\st{\sin{\theta}}
\def\cb{\cos\beta}
\def\sb{\sin{\beta}}
\def\cdb{\cos^2{\beta}}
\def\sdb{\sin^2{\beta}}
\def\sidb{\sin{2\beta}}
\def\cg{\cos{\gamma}}
\def\sg{\sin{\gamma}}
\def\sidg{\sin2\gamma}
\def\codg{\cos{2\gamma}}
\def\cp{\cos{\psi}}
\def\sp{\sin{\psi}}
\def\cdp{\cos^2{\psi}}
\def\sidp{\sin{2\psi}}
\def\DP{\Delta \ptau}
\def\br{ {\bf BR}}
\def\s2thw{\sin^{2}\theta^{\mbox{\scriptsize lept}}_{\mbox{\scriptsize eff}}}
\def\kos{K^{0}_{\! \mathrm S}}
\def\kol{K^{0}_{\! \mathrm L}}
\newcommand{\Z}{\mbox{\rm Z}}
\newcommand{\MZ}{M_{\rm Z}}
\newcommand{\nut}{\nu_{\tau}}
\newcommand{\num}{\nu_{\mu}}
\newcommand{\nue}{\nu_{\elec}}
\newcommand{\Tauto}{\tau^- \rightarrow}
\def\pio{\pi^0}
\def\ko{K^0}
\def\nt{\nu_\tau}
\def\tenn{\tau^- \rightarrow \rm{e}^- \bar{\nu_e}\nu_\tau}
\def\tlnn{\tau^- \rightarrow l^- \bar{\nu_e}\nu_\tau}
\def\tmnn{\tau^- \rightarrow \mu^- \bar{\nu_{\mu}}\nu_\tau}
\def\tpn{\tau^- \rightarrow \pi^- \nu_\tau}
\def\thn{\tau^- \rightarrow h^- \nu_\tau}
\def\tpno{\tau^- \rightarrow \pi^- \pi^{0} \nu_\tau}
\def\tpnoo{\tau^- \rightarrow \pi^- \pi^0 \pi^0 \nu_\tau}
\def\tpngoo{\tau^- \rightarrow \pi^- \geq 2 \pio \nu_\tau}
\def\tpnooo{\tau^- \rightarrow \pi^- \geq 3 \pio \nu_\tau}
\def\tpppn{\tau^- \rightarrow \pi^-\pi^-\pi^+  \nu_\tau}
\def\tpppno{\tau^- \rightarrow \pi^- \pi^-\pi^+ \geq 1 \pi^0\nu_\tau}
\def\tpkn{\tau^- \rightarrow \pi^-(K^-)\nu_\tau}
\def\trn{\tau^- \rightarrow \rho^- \nu_\tau}
\def\tan{\tau^- \rightarrow {a_1}^- \nu_\tau}
\def\thrn{\tau^- \rightarrow h^- \pio \nu_\tau}
\def\than{\tau^- \rightarrow h^- \pio \pio\nu_\tau}
\def\thgan{\tau^- \rightarrow h^- \geq\! 2 \pio\nu_\tau}
\def\thgann{\tau^- \rightarrow h^- \geq\! 3 \pio\nu_\tau}
\def\tth{\tau^- \rightarrow h^-h^+h^- \nu_\tau}
\def\thp{h^-h^+h^-\pio\nu_\tau}
\def\tthp{\tau^- \rightarrow \thp}
\def\tthpp{\tau^- \rightarrow h^-h^+h^- \geq\! 2 \pio\nu_\tau}
\def\ttp{\tau^- \rightarrow h^-h^+h^- \geq 0~\gamma \nu_\tau}
\def\taha{\tau^- \rightarrow hadrons~\nu_\tau}
\newcommand{\TP}{\Tauto \pi^- \nut}
\newcommand{\TK}{\Tauto K^-\nut}
\newcommand{\HH}{h^-\nut}
\newcommand{\HHK}{h^-\nut(\kol)}
\newcommand{\TH}{\Tauto\HH}
\newcommand{\THK}{\Tauto\HHK}
\newcommand{\MU}{\mu^- \bar{\nut} \num}
\newcommand{\TMU}{\Tauto\MU}
\newcommand{\EL}{\elec^- \bar{\nue} \nut}
\newcommand{\TEL}{\Tauto\EL}
\newcommand{\TRO}{\Tauto \rho^- \nut}
\newcommand{\PZ}{\pi^-\pi^0 \nut}
\newcommand{\PK}{\pi^- K^0 \nut}
\newcommand{\KZ}{K^-\pi^0 \nut}
\newcommand{\HZ}{h^-\pi^0 \nut}
\newcommand{\TPZ}{\Tauto \PZ}
\newcommand{\TKZ}{\Tauto \KZ}
\newcommand{\THZ}{\Tauto \HZ}
\newcommand{\PZZ}{\pi^-\pi^0\pi^0 \nut}
\newcommand{\HZZ}{h^-\pi^0\pi^0 \nut}
\newcommand{\TPZZ}{\Tauto \PZZ}
\newcommand{\THZZ}{\Tauto \HZZ}
\newcommand{\HGZZ}{h^-\geq \! 2\pi^0 \nut}
\newcommand{\THGZZ}{\Tauto h^-\geq\! 2\pi^0 \nut}
\newcommand{\PZZZ}{\pi^-3\pi^0 \nut}
\newcommand{\HZZZ}{h^-3\pi^0 \nut}
\newcommand{\TPZZZ}{\Tauto \PZZZ}
\newcommand{\THZZZ}{\Tauto \HZZZ}
\newcommand{\HGZZZ}{h^-\geq \! 3\pi^0 \nut}
\newcommand{\THGZZZ}{\Tauto h^-\geq\! 3\pi^0 \nut}
\newcommand{\PPP}{2\pi^-\pi^+ \nut}
\newcommand{\HHH}{2h^-h^+ \nut}
\newcommand{\TPPP}{\Tauto \PPP}
\newcommand{\THHH}{\Tauto \HHH}
\newcommand{\PPPZ}{2\pi^-\pi^+\pi^0 \nut}
\newcommand{\HHHZ}{2h^-h^+\pi^0 \nut}
\newcommand{\HHHGZ}{2h^-h^+\geq\!1\pi^0 \nut}
\newcommand{\TPPPZ}{\Tauto \PPPZ}
\newcommand{\THHHZ}{\Tauto \HHHZ}
\newcommand{\THHHGZ}{\Tauto \HHHGZ}
\newcommand{\PPPPP}{3\pi^-2\pi^+ \nut}
\newcommand{\HHHHH}{3h^-2h^+ \nut}
\newcommand{\PPPZZ}{3\pi^\pm 2\pi^0 \nut}
\newcommand{\PPPZZZ}{3\pi^\pm 3\pi^0 \nut}
\newcommand{\PZZZZ}{\pi^- 4\pi^0 \nut}
\newcommand{\PPPPPZ}{5\pi^\pm \pi^0 \nut}
\newcommand{\HHHHHZ}{5h^\pm \pi^0 \nut}
\newcommand{\TAONE}{\Tauto \aone^- \nut}
\newcommand{\TKPP}{\Tauto K^- \pi^- \pi^+ \nut}
\newcommand{\TKKP}{\Tauto K^- K^+ \pi \nut}
\newcommand{\TPKO}{\Tauto \pi^- K^0 \nut X}
\newcommand{\TKKO}{\Tauto K^- K^0 \nut X}
\newcommand{\THKO}{\Tauto h^- K^0 \nut X}
\newcommand{\TPKN}{\Tauto (K/\pi)^- n\pi^0 \nut}
\newcommand{\HN}{h^- n\pi^0 \nut}
\newcommand{\THN}{\Tauto \HN}
\newcommand{\TPN}{\Tauto \pi^- n\pi^0 \nut}
\newcommand{\TKN}{\Tauto K^- n\pi^0 \nut}
\newcommand{\TPPPNG}{\Tauto \pi^-\pi^-\pi^+ (n \gamma) \nut}
\newcommand{\TPNG}{\Tauto \pi^-\pi^-\pi^+ (n \gamma) \nut, n \ge 0}
\newcommand{\K}{\phantom{-}}
\newcommand{\Eovp}{E_{ass}/\pnoOD}
\newcommand{\etal}{{\em et al.}\/}
\newcommand{\barve}{\bar{v}_{\mbox{\footnotesize e}}}
\newcommand{\barae}{\bar{a}_{\mbox{\footnotesize e}}}
\newcommand{\barvt}{\bar{v}_{\tau}}
\newcommand{\barat}{\bar{a}_{\tau}}
\newcommand{\barvl}{\bar{v}_l}
\newcommand{\baral}{\bar{a}_l}
\renewcommand{\topfraction}{0.9}
\renewcommand{\bottomfraction}{0.9}
\renewcommand{\textfraction}{0.1}
\setcounter{totalnumber}{1}
\setcounter{topnumber}{1}
\setcounter{dbltopnumber}{1}
\setcounter{bottomnumber}{1}
\section{Introduction}
\label{sec:introduction} 
Historically, there  has been an inconsistency in  the measured $\tau$
exclusive  branching ratios: the  measurements of  exclusive branching
ratios to decay  modes containing one charged particle  did not sum up
to the inclusive branching ratio for a charged multiplicity of one.  A
number  of analyses  have attempted  and succeeded  in  resolving this
question~\cite{pdg96,cellobr,alephbr1,alephbr2}.   However the current
world  average  values from  direct  measurements  of the  topological
branching  ratios in the  Particle Data  Group listings~\cite{pdg2000}
have  uncertainties which  are  significantly larger  than the  values
obtained through  combined fits to all  the $\tau$ decay  data. In the
case  of the 1-prong  branching ratio  ($\Gamma_2$ in~\cite{pdg2000}),
the difference between the average and  the fit value is more than two
standard deviations.

This paper presents a  dedicated simultaneous measurement of the decay
rates of the $\tau$ lepton to  different final states as a function of
the charged particle multiplicity.

The relevant  components of the  DELPHI detector and the  data-set are
described  in  Section~\ref{sec:detector}.    The  definition  of  the
measured  branching ratios  and the  method  used to  derive them  are
introduced          in          Section~\ref{sec:method}.           In
Section~\ref{sec:tautauselection}  the preselection  of the  sample of
$\eett$ events is described. Section~\ref{sec:reconstruction} contains
a description of  the reconstruction of charged particles,  as well as
the reconstruction  algorithms for photons and  hadrons which interact
with  the detector  material  before the  tracking subdetectors.   The
$\tau$  decays  are classified  according  to  their charged  particle
multiplicity.  This  is described in  Section~\ref{sec:topology}.  The
results     and     systematic     studies    are     presented     in
Sections~\ref{sec:fitmethod} and~\ref{sec:systematics} respectively.

\section{Experimental apparatus and data sample}
\label{sec:detector} 
The  DELPHI  detector and  its  performance  are  described in  detail
elsewhere~\cite{detect,delphi_performance}.    The  subdetector  units
particularly relevant  for this  analysis are summarised  here.  These
detector  components covered the  full solid  angle considered  in the
analysis  except where specified,  and sat  in a  1.2~Tesla solenoidal
magnetic field parallel to the z-axis\footnote{In the DELPHI reference
frame the origin is at the centre of the detector, coincident with the
interaction  region, the  z-axis is  parallel to  the e$^-$  beam, the
x-axis points horizontally towards the  centre of the LEP ring and the
y-axis  is vertically  upwards.   The co-ordinates  r,$\phi$,z form  a
cylindrical coordinate system, while  $\theta$ is the polar angle with
respect to the z-axis.}.

The reconstruction of charged particles in the barrel region of DELPHI
used  a  combination  of  the  measurements  in  four
different  cylindrical subdetectors: a  silicon Vertex  Detector (VD);
the Inner  Detector (ID),  consisting of a  jet chamber tracker  and a
wire chamber  used for trigger  purposes; the Time  Projection Chamber
(TPC); the Outer Detector (OD).

The VD  had three layers of  silicon micro-strip modules,  at radii of
6.3,~9.0 and~11.0~cm  from the beam  axis.  The space  point precision
was  about  $8~\mu$m  in~r-$\phi$.  For  data  from  1993  onwards,  a
measurement  of  r-z  was  produced  in the  outermost  and  innermost
layers.  This  had a  precision  of  about  15~$\mu$m.  The-two  track
resolution was 100~$\mu$m~in~r-$\phi$ and 200~$\mu$m in r-z.

The ID had an inner radius of  12~cm and an outer radius of 28~cm. The
inner jet chamber part lay at radii below 22~cm.
It had a two-track  resolution in r-$\phi$ of
1~mm and a precision in  r-$\phi$ of $50~\mu$m. The outer triggering
part was a five-layer wire chamber. This
was replaced for  the 1995 data with a  straw detector containing less
material.

The  TPC, extending  from  30~cm to  122~cm  in radius,  was the  main
detector  for   charged  particle  reconstruction.    The  main  track
reconstruction information  was provided  by 16 concentric  circles of
pads  which supplied  up to  16  three-dimensional space  points on  a
track. In  addition, ionisation information  was extracted from  up to
192 wires. This was  used for particle identification purposes.  Every
$60^\circ$  in $\phi$  there was  a boundary  region  between read-out
sectors  about  $1^\circ$  wide  which  had  no  instrumentation.   At
$\qcth=0$ there  was a cathode  plane which caused a  reduced tracking
efficiency in the polar  angle range $|\qcth|\!<\!0.035$.  
The TPC had
a two-track resolution of about 3~mm in r-$\phi$ and~1.5~cm in~z.

The OD, with~5~layers of drift cells  at a radius of 2~m from the beam
axis, was  important  for  the  momentum  determination  of  energetic
particles.

The principal  device for electron  and photon identification  was the
High density  Projection Chamber (HPC), located outside  the OD, which
allowed full reconstruction of  the longitudinal and transverse shower
components. It covered the  polar angle region $|\qcth|\!<\!0.75$.  In
the forward region, $0.800\!<\!|\qcth|\!<\!0.985$, 
the calorimetry was performed
by   a    lead-glass   array.    For   very   small    polar   angles,
the  calorimetry was performed  by the luminosity
monitors, the SAT in 1992 and 1993,  and the STIC in 1994 and 1995. In
the region $0.75\!<|\qcth|\!<\!0.8$, there was no electromagnetic
calorimeter.

Outside  the  magnet solenoid  lay  the  hadron  calorimeter and  muon
chambers, which were used for hadron and muon identification.

The  data  were  collected  by   the  DELPHI  detector  from  the  LEP
electron-positron  collider,  in  the  years  1992  through  1995,  at
centre-of-mass energies $\sqrt{s}$ of  the $\ee$ system between 89 and
93~GeV, on or near to the  $Z$ resonance. It was required that the VD,
ID, TPC and HPC were in good operational condition for the data sample
analysed.   The   integrated  luminosity   of  the  data   sample  was
135~pb$^{-1}$ of which about  100~pb$^{-1}$ was taken at $\sqrt{s}$ of
91.3~GeV, close to  the maximum of the $Z$  boson resonance production
cross-section.
        
The selection procedures were studied with samples of simulated events
which had  been passed through  a detailed simulation of  the detector
response~\cite{delphi_performance}  and  reconstructed  with the  same
program  as the  real data.   
Separate samples were produced corresponding to the detector
configurations in different years.
The  Monte Carlo  event generators  used
included:     KORALZ~4.0~\cite{jadach}     for     $\eett$     events;
DYMU3~\cite{dymu3}   for  $\eemm$  events;   BABAMC~\cite{babamc}  and
BHWIDE~\cite{bhwide}  for $\bhab$ events;  PYTHIA 5.7~\cite{sjostrand}
for $\eeha$ events.  Four-fermion final states were produced using two
different generators.  BDK~\cite{berends}  was used for reactions with
four leptons in the final state. This included two-photon events where
one or two  e$^+$ or e$^-$ were not observed  in the detector.  
TWOGAM~\cite{twogam}
was  used   to  generate  $\ee\to(\ee)\qq$   events.   The  KORALZ~4.0
generator   incorporated  the  TAUOLA~2.5~\cite{tauola}   package  for
modelling  the  $\tau$ decays.  A  value of  1.2\%  was  used for  the
branching ratio of the Dalitz decay $\pi^0\to\gamma\ee$.

\section{Method}
\label{sec:method}
From  a high  purity sample  of $Z\to\tt$  events, $\tau$  decays were
identified with charged multiplicity one, three or five, corresponding
to the topological branching ratios defined~as:
\[\begin{array}{lll}
\Bone  & \equiv B(\Tauto  {\mathrm   (particle)}^-  
                      \!\geq\!0\pi^0\!\geq\!0K^0\nut(\bar{\nu}));\\ 
\Bthree &\equiv B(\Tauto 2h^-h^+ \!\geq\!0\pi^0\!\geq\!0K^0\nut);\\ 
\Bfive & \equiv B(\Tauto 3h^-2h^+ \!\geq\!0\pi^0\!\geq\!0K^0\nut),
\end{array}\]
where $h$ is either a charged $\pi$ or $K$ meson.  In this definition,
the  $\kos$ meson  was treated  as a  neutral particle,  even  when it
decayed  to $\pi^+\pi^-$  before  the tracking  detectors.  In  $\tau$
decays  only about  5\% of  $\kos$ mesons  decay before  the  VD.  The
$h^-K^0(neutrals)\nut$ component  was therefore treated  as signal for
the  1-prong  $\tau$ decays  and  background  for  the 3-prong  $\tau$
decays.   Although a $\pi^0$  meson which  decayed to  the $\gamma\ee$
final state  produced two charged particles observed  in the detector,
it  was also  treated  as  a neutral  particle,  following the  $\tau$
branching    ratio   definition   given    by   the    Particle   Data
Group~\cite{pdg2000}.   Thus the  branching ratios  $\Bone$, $\Bthree$
and $\Bfive$ correspond  to the parameters $\Gamma_{2}$, $\Gamma_{55}$
and~$\Gamma_{101}$ respectively~in~\cite{pdg2000}.

No attempt was made to measure $\tau$ decays to more than five charged
particles  and  it was  assumed  that  such  decays had  a  negligible
branching ratio.   A 90\%  CL upper limit  on the  inclusive branching
ratio for the $\tau$ decay to  seven charged particles has been set at
$2.4\times10^{-6}$ by the CLEO experiment~\cite{cleosevenlimit}.

At  LEP it is  possible to  separate cleanly  $\tt$ events  from other
final states, permitting the efficient selection of a 
high purity $\tau$ sample.
One can  measure the branching ratio  $B(\tau\to X)$ for  the decay of
the $\tau$ to a final state $X$ using the expression
\begin{equation}
\label{eqn:eq3}
  B(\tau\to X) = \frac{N_X} {N_{\tau}} \cdot \frac{1-b_X} {1-b_{\tau}}
                                    \cdot       \frac{\epsilon_{\tau}}
                                     {\epsilon_X } ,
\end{equation}
where $N_X$  is the number of  identified decays of type  $X$ found in
the  sample of  $N_{\tau}$ $\tau$-decay  candidates,  preselected with
efficiency $\epsilon_{\tau}$ with a background fraction of $b_{\tau}$.
$\epsilon_X$  is  the  total (preselection  $\times$  identification)
efficiency for selecting the decay mode $\tau\to X$, with a background
fraction of $b_X$, including background from other $\tau$ decay modes.
In a $\tt$ event selection  without specific requirements on either of
the  two $\tau$  candidates in  the event,  $\epsilon_{\tau}$  will be
identical to the $\tt$ event selection efficiency.

In   the   case   where   several  branching   ratios   are   measured
simultaneously,  candidate  $\tau$  decays  can  be  classified  as  a
function of the detector components used in the charged particle track
reconstruction.
These  classes (detailed in
Section~\ref{sec:topology}) typically  have different signal  to 
background
ratios for different types of decays and thus contain different levels
of  information. To obtain  the branching  ratios, a  fit can  then be
performed  to   the  predicted   number  $N_{i,pred}$  of   decays  in
a~class~$i$:
\begin{equation}
\label{eqn:classpred}
N_{i,pred} = N_\tau (\epsilon_{i,1}\Bone
           + \epsilon_{i,3}\Bthree
           + \epsilon_{i,5}\Bfive )
           + N_{i,bkg}~,
\end{equation}
where $\epsilon_{i,m}$  is the probability  of a $\tau$ decay  of true
multiplicity $m$ being attributed to class $i$
and  $N_{i,bkg}$ is  the  estimated  background in  class  $i$ due  to
non-$\tt$ events.  This can be extended to a simultaneous fit for both
$\tau$  decays  in a  candidate  $\tt$  event.   The predicted  number
$N_{i,j,pred}$ of $\tt$ events  with one $\tau$~decay in class~$i$ and
the other in class~$j$~is
\begin{equation}
\label{eqn:classpred2}
N_{i,j,pred} = N_{\tau\tau} \sum_{m,n}  
             \epsilon_{ij}^{mn} B_m B_n + N_{i,j,bkg}~,
\end{equation}
where $N_{i,j,bkg}$  is the  estimated number of  non-$\tt$ background
events in  class $i,j$ and $\epsilon_{ij}^{mn}$ is  the probability of
an event with decays of true multiplicity $m$ and $n$ being attributed
to class $i,j$.  $N_{\tau\tau}$ is the number of  $\tt$ events and can
be left free  and obtained from a fit to  the different classes.  This
method takes into account correlations between the two observed $\tau$
decays in  an event and  has the advantage that  non-$\tt$ backgrounds
tend to populate event classes which have correlations between the two
candidate $\tau$  decays (e.g. dimuons in the  1-versus-1 topology but
not   in   1-versus-3)  and   thus   exhibits   less  sensitivity   to
backgrounds.  It was  used as  the principal  method in  the following
analysis.

\boldmath
\section{$\eett$ preselection}
\label{sec:tautauselection}
\unboldmath In $\eeztt$  events at $\sqrt{s} = M_Z$,  the $\tau^+$ and
$\tau^-$ are produced  back-to-back, ignoring radiative effects.  Each
$\tau$  decays, producing  one, three,  or more  charged  particles in
addition to  one or two  neutrinos and, possibly, neutral  mesons. All
particles apart from the neutrinos can be detected in DELPHI.  A $\tt$
event is thus characterised by  two low multiplicity jets which appear
approximately  back-to-back  in  the  laboratory.   The  loss  of  the
neutrinos implies that not all the  energy in the event is seen in the
detector.

Except  where  explicitly  stated,  all  quantities  calculated  using
charged particles used only charged  particle tracks which had hits in
at least  the TPC or OD,  to ensure good  momentum reconstruction from
the long track length, and had an impact parameter 
with respect to the centre of the interaction region
of less than 1.5~cm
in~r-$\phi$ and 4.5~cm in~z.  The basic $\eett$ preselection described
below  has some  differences  compared with  that  used previously  in
studies       of       the       $\tau$       leptonic       branching
ratios~\cite{delphileptonicbr99}.  There are slightly modified cuts to
reduce non-$\tau$ backgrounds and the addition of subsamples of events
containing $\tau$ decays with  a nuclear reinteraction in the detector
material   and  $\tau$   decays   where  one   charged  particle   was
reconstructed only in the VD and ID tracking subdetectors.

\boldmath
\subsection{Preselection criteria}
\label{sec:preselectioncriteria}
\unboldmath  To ensure  that the  $\tau$  decay products  were in  the
region of DELPHI corresponding to  the HPC acceptance, the thrust axis
of  the event,  calculated  using only  charged  particle tracks,  was
required to lie  in the polar angle region  defined by $|\cos\theta| <
0.732$.  The event was split  into two hemispheres, each associated to
a candidate  $\tau$ decay,  by the plane  perpendicular to  the thrust
axis and containing the point at the centre of the interaction region.
Each hemisphere had to contain  at least one charged particle.  It was
required  that at least  one charged particle  lie in  the polar  angle region
defined by  $|\cos\theta| > 0.035$ in  order to reduce  the effects of
the TPC tracking inefficiency near $\cos\theta=0$.

Most hadronic  $Z$ decays  were rejected by  requiring that  the event
contain  at most  eight charged  particle tracks  with hits  in either
the~TPC~or~OD and satisfying the impact parameter cuts described above.

Four-fermion  events  were  suppressed  by requiring  that  the  event
isolation angle be  greater than $160^\circ$. This was  defined as the
minimum  angle  between  any  pair  of charged  particles  in
opposite $\tau$ decay hemispheres.

Backgrounds from  $\mm$ and  $\ee$ final states  and cosmic  rays were
reduced  by   requiring  that  the   isolation  angle  be   less  than
$179.5^\circ$ for events with only two charged particles.

\begin{figure}[t]
  \begin{center}
  \vspace{-25pt}
\epsfig{figure=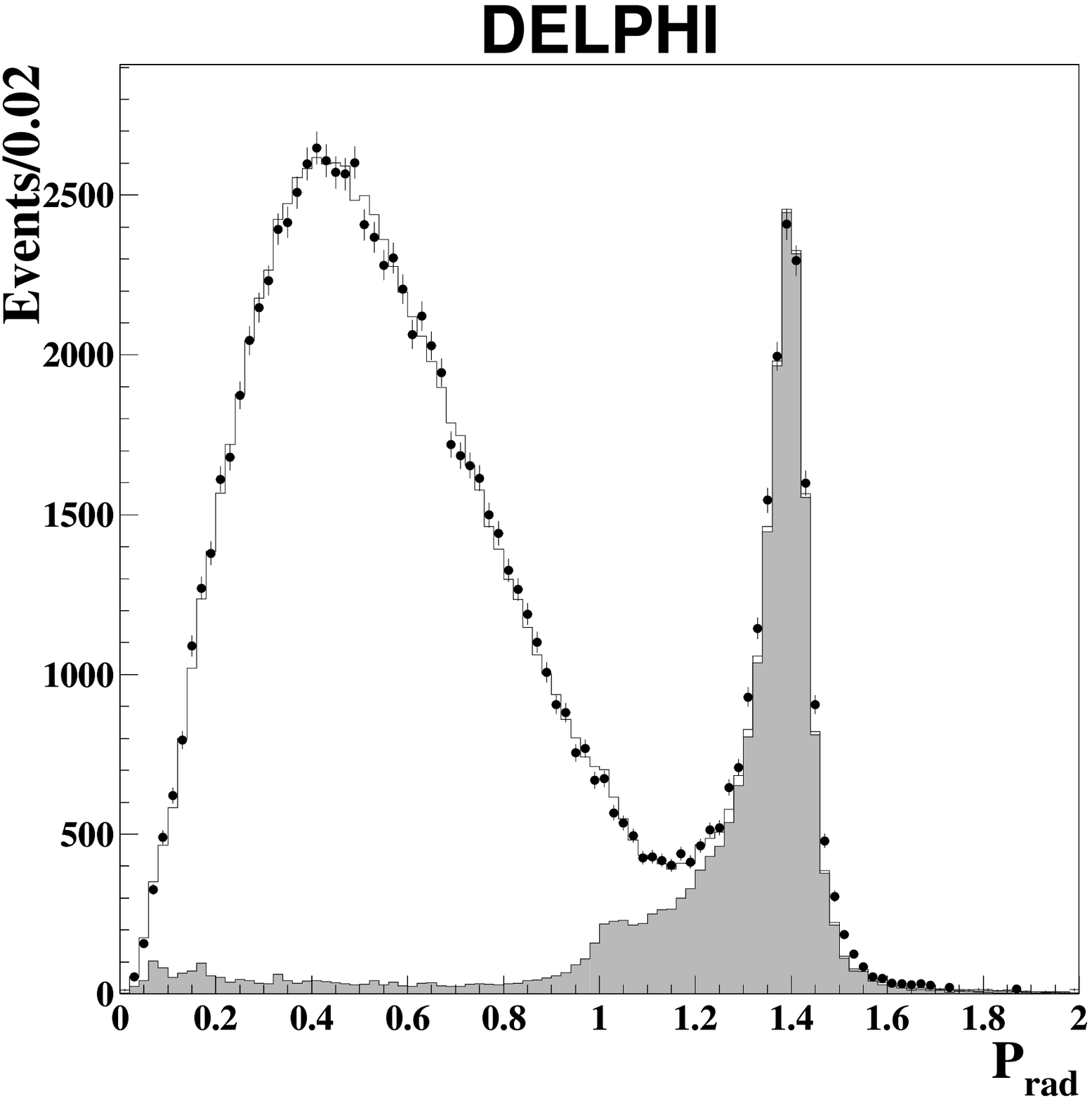,width=0.49\linewidth}
\put(-50,170){\bf\huge a}                    
\hfill
\epsfig{figure=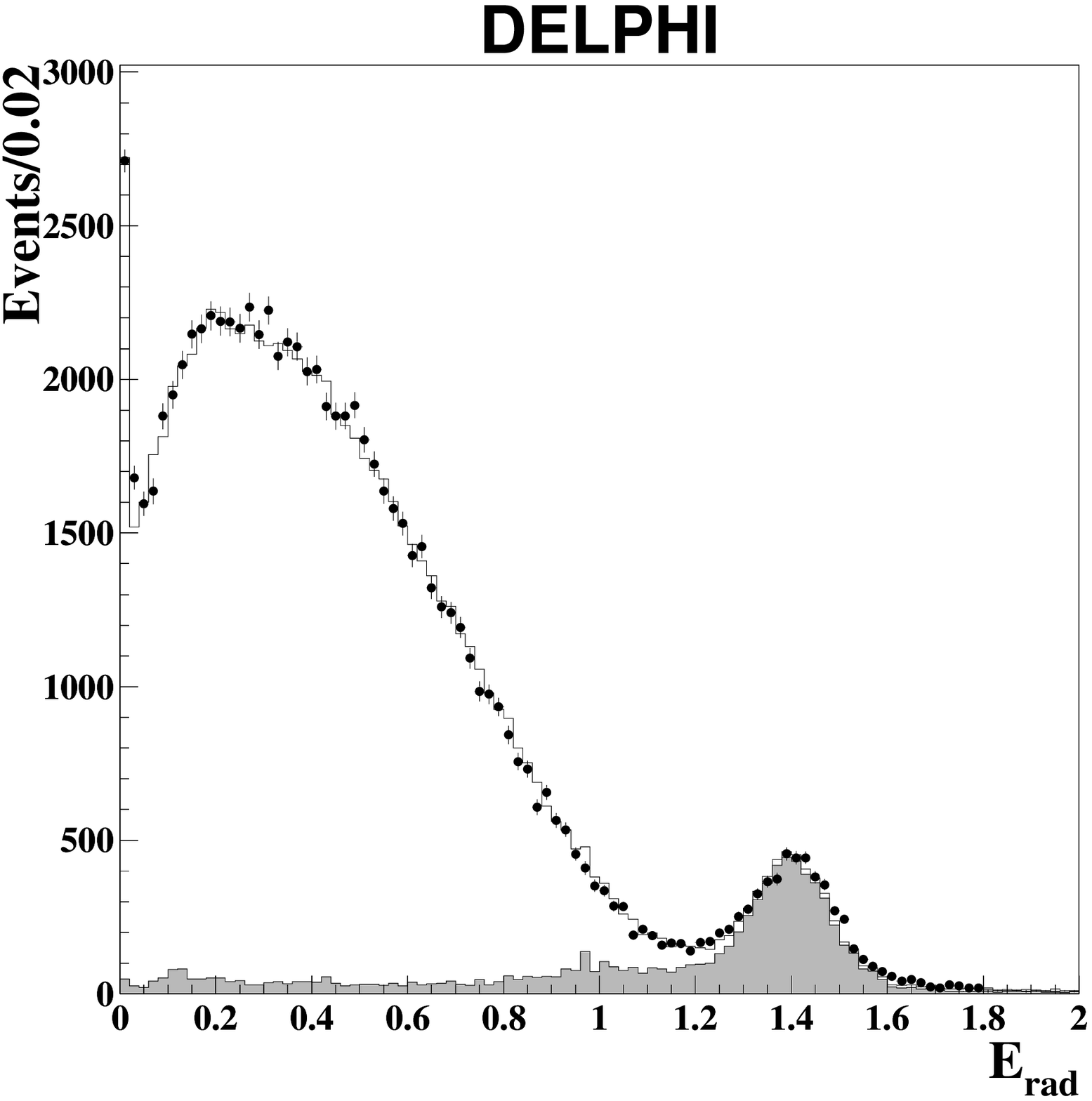,width=0.49\linewidth}
\put(-50,170){\bf\huge b}
    \caption{\it Distributions  of a) $p_{rad}$ and b)~$E_{rad}$ 
     variables used  in $\tt$ selection.  The dots are data, the line is
     simulation.  The shaded  area is  non-$\tt$ background.   In both
     cases  the  selected  sample  lies  in the  region  below  a  cut
     at~1.0. The spike  in the first bin of plot~(b)  is due to events
     where both  $\tau$'s decay to either  a muon or  a charged hadron
     which does  not interact  in the HPC.   This bin has  been scaled
     down by a factor two for presentational purposes.}
    \label{fig:prad}
  \end{center}
\end{figure}

The  $\mm$ and $\ee$  contamination was  reduced further  by requiring
that           both           $p_{rad}=(\frac{|\vec{p}_1|^2}{p_{1}'^2}
+\frac{|\vec{p}_2|^2}{p_{2}'^2})^{^{1/2}}$                         and
$E_{rad}=(\frac{E_1^2}{E_{1}'^2}+ \frac{E_2^2}{E_{2}'^2})^{^{1/2}}$ be
less than  unity.  The variables  $\vec{p}_1$ and $\vec{p}_2$  are the
momenta  of  the highest  momentum  charged  particles in  hemispheres
1~and~2  respectively.  The  quantity $p_{1}'$  was obtained  from the
formula
\begin{equation}         
p_{1}'=\sqrt{s}\sin\theta_2/(\sin\theta_1+\sin\theta_2
+|\sin(\theta_1+\theta_2)|)~~~,
\label{eqn:pprime}
\end{equation}
and   $p_{2}'$  by  analogy   with  the
indices~1~and~2  interchanged.  The  angles $\theta_1$  and $\theta_2$
are  the polar  angles of  the  highest momentum  charged particle  in
hemispheres 1~and~2  respectively.  The variables $E_1$  and $E_2$ are
the total  electromagnetic energies  deposited in cones  of half-angle
$30^\circ$  about  the momentum  vectors  $\vec{p}_1$ and  $\vec{p}_2$
respectively, while  $E_{j}'=cp_{j}'$,~for $j=1,2$.  These definitions
of $p_{rad}$ and $E_{rad}$ are identical to those used for the $\eett$
measurements  at LEP-2~\cite{lep2xsec}.  They  take into  account more
correctly the radiative events where  the radiative photon is close to
the  beam   axis  than  the   definition  used  previously   in  LEP-1
analyses~\cite{delphileptonicbr99,lineshape}.   In  the  limit  of  no
radiation   the  new   and  old   definitions  are   equivalent.   The
distributions    of   $p_{rad}$   and    $E_{rad}$   are    shown   in
Fig.~\ref{fig:prad}a and  Fig.~\ref{fig:prad}b respectively, after all
cuts have been imposed except the one on the displayed variable.
There are some small data/simulation discrepancies in the momentum
resolution, visible in the dimuon peak at $p_{rad}\approx1.4$. This
had a negligible effect on the analysis.

Much  of the remaining  background from  the dileptonic  channels came
from events containing hard radiation  lying far from the beam.  These
events  should lie  in a  plane.  Where  two charged  particles  and a
photon  were visible  in the  detector,  such events  were removed  by
requiring that the  sum of the angles between  the three particles was
greater than $359.8^\circ$.

A  further  reduction  in  four-fermion  background  was  achieved  by
requiring a minimum  visible energy in the event.   The visible energy
$E_{vis}$ is the sum of the  energies of all charged particles and the
electromagnetic energies  of all neutral  particles, neglecting energy
deposits recorded by  the very forward calorimeters (SAT  and STIC) at
angles  of less  than  $12^\circ$  from the  beam  axis.  Events  were
accepted  if  $E_{vis}$ was  greater  than $0.09\times\sqrt{s}$.   For
events with  only two charged  particles, an additional  condition was
applied that the vectorial sum  of the charged particle momenta in the
r-$\phi$  plane  be  greater  than  0.4~GeV/$c$.  Four-fermion  events
typically have very low values of of this quantity compared with $\tt$
events.

Most  cosmic  rays  were  rejected  by  the  upper  cut  on  isolation
angle. Further  rejection was  carried out by  requiring at  least one
charged particle in the event have an impact parameter with respect to
the interaction region  of less than~0.3~cm in the  r-$\phi$ plane and
that in the $\tau$ decay hemisphere opposite to this particle at least
one charged particle have an impact parameter in the r-$\phi$ plane of
less than~1.5~cm.   It was also  required that both  event hemispheres
have a charged  particle track whose perigee point lay  within 4.5~cm of the
interaction region in~z.

Additional  $\tt$  events were  selected  where  one  of the  $\tau$'s
produced  a 1-prong  decay and  the  other $\tau$  decay produced  one
charged hadron which interacted  with the detector material inside the
TPC  and  where  the  interaction  was reconstructed  by  the  nuclear
reinteraction      reconstruction      algorithm     described      in
Section~\ref{sec:nucints}.  The cuts in  $p_{rad}$ and  $E_{rad}$ were
not applied. This increased the selection efficiency by about 1\% with
a relative background of~0.8\%.

A further class of events was  selected where one of the $\tau$ decays
left one  charged particle track with hits  in only the VD  and the ID
due  to  inefficiencies in  the  TPC and  OD.   Many  of these  tracks
contained no polar angle information and had poor momentum resolution,
so they were excluded from the calculation of event quantities such as
the thrust~or~$p_{rad}$.  The~$p_{rad}$  requirement was replaced with
a cut on the momentum of  the highest momentum charged particle in the
opposite hemisphere,  requiring it  to be less  than 90\% of  the beam
momentum. This  cut was tightened to  70\% if there  was an identified
muon in the hemisphere.  This subsample of events contributed an extra
4\% of events with a relative background of 5\%.

The total  efficiency, including all  classes of events, was estimated
from  simulation to  be  $(51.74\pm0.04)\%$.  This  corresponds to  an
efficiency of approximately 85\% within the angular acceptance cuts.

\boldmath
\subsection{Backgrounds}
\label{sec:backgrounds}
\unboldmath  
\begin{figure}[t]
  \begin{center}
   \vspace{-15pt}
   \epsfig{figure=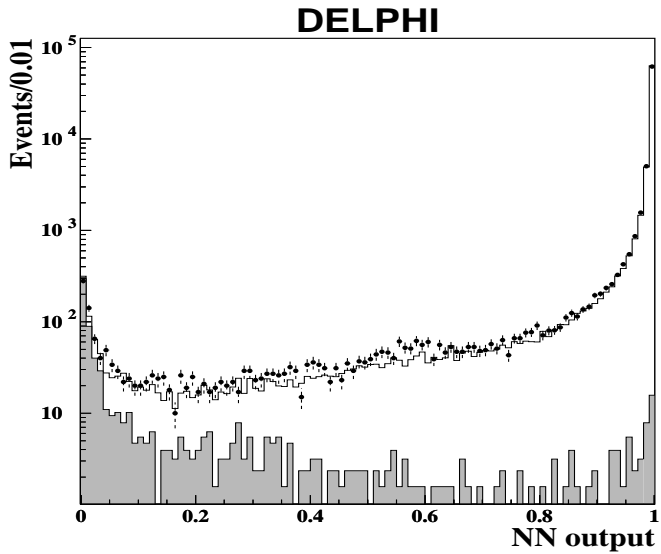,width=0.6\linewidth,height=8cm}
    \caption{\it The output neuron
distribution  for the  neural  network used  for  the final  ${\mathit
q\!\bar{q}}$  rejection.   The  dots  are  data,  the    line  is
simulation  and  the  shaded  region  shows  the  simulated  ${\mathit
q\!\bar{q}}$ contribution. The selected  events lie in the region with
output neuron values above 0.05.}
    \label{fig:nnqq}
  \end{center}
\end{figure}
Knowledge  of the  background  from  $\qq$  events is  of
particular  importance in this  analysis as  it constitutes  a serious
background  in the higher  multiplicity topologies.   Uncertainties in
the low multiplicity backgrounds have a reduced effect on the measured
branching  ratios due  to  the  large $B_1$  value  and the  resultant
cancellation  in Eqn.~\ref{eqn:eq3}.  After the cuts  described above,
the $\qq$  background estimated from  simulation is $(0.80\pm0.05)\%$,
where the  error is purely statistical.  To  reduce systematic effects
due to $\qq$  background, a further level of  rejection was performed.
A  number of observables  display separation  power between  $\tt$ and
$\qq$  events  while  being  only  weakly dependent  on  the  topology
classification.  These  were: the  isolation angle; the  event thrust;
the  number  of  neutral  electromagnetic  showers in  an  event;  the
invariant  masses  reconstructed   using  charged  particles  in  each
candidate $\tau$ decay hemisphere of an event, and the invariant masses
reconstructed  using  both  charged  and  neutral  particles  in  each
candidate  $\tau$  decay hemisphere.   These  variables  were used  as
inputs to a neural network with  one hidden layer of eight neurons and
a single  output neuron.  The  output neuron distribution is  shown in
Fig.~\ref{fig:nnqq}. 
A cut  at  0.05  in the  output  neuron gave  a
rejection factor of about three, leaving a background of $0.29\%$ from
$\eeha$,  while  rejecting only  $1.3\times10^{-3}$  of the  remaining
$\tt$ events. Discrepancies between data and simulation in this
variable are discussed in Section~\ref{sec:systematics_preselection}.

Background levels  were estimated as accurately as  possible using the
data.   Variables sensitive to  a particular  type of  background were
chosen.   The background  levels  were extracted  by  fitting for  the
background  contribution in  these variables  assuming  the background
shape from  the simulation.  Typically  a relative precision  of order
10\% or better was achieved, ensuring systematic uncertainties  on the
topological branching ratios below the expected statistical precision.

The background from $\eeha$ was estimated using the data by performing
a fit  to the distribution of  the neural network  output neuron.  The
shape of the background distribution was taken from simulation and the
fit   was  performed   for   the  normalisation   factor.   A   factor
$0.92\pm0.03$ was  obtained compared to the simulation. 

\begin{figure}[t]
  \begin{center}
   \vspace{-10pt}
   \epsfig{figure=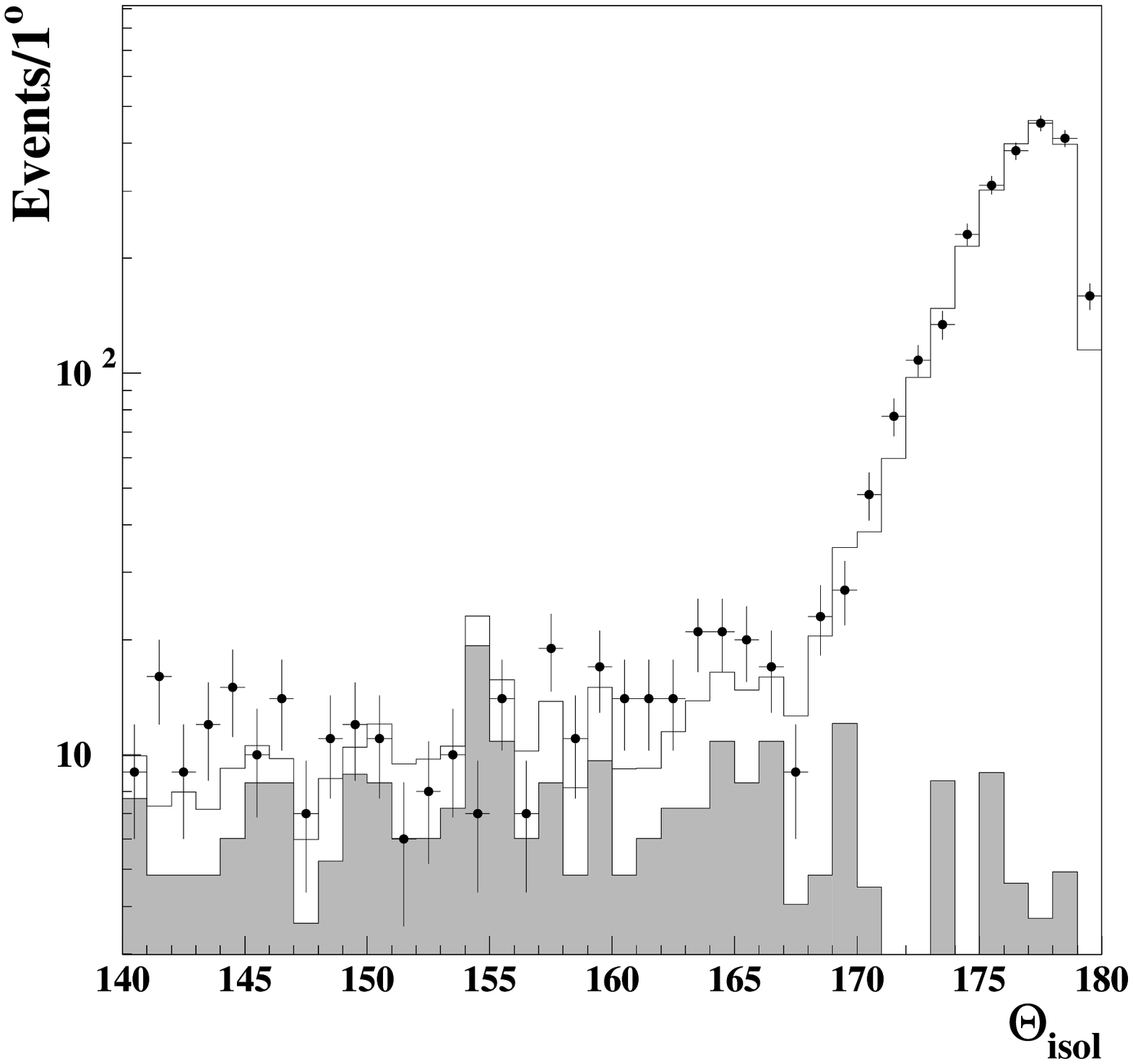,width=0.33\linewidth,height=7cm,bbllx=0,bblly=0,bburx=515,bbury=530}
\put(-120,160){\bf\huge a}    
\hfill     
    \epsfig{figure=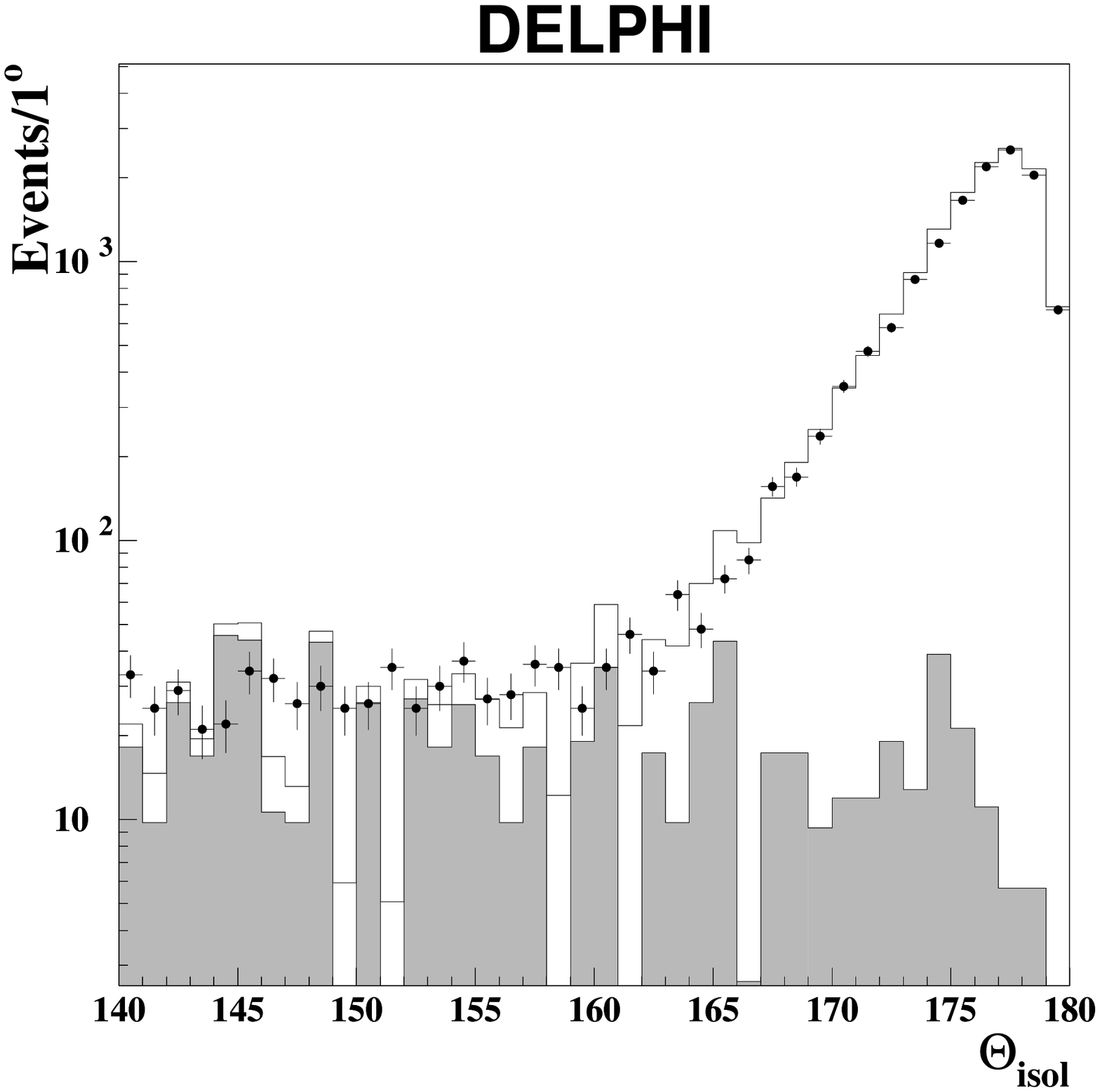,width=0.33\linewidth,height=7cm,bbllx=0,bblly=0,bburx=515,bbury=530}
\put(-120,160){\bf\huge b}    
\hfill     
    \epsfig{figure=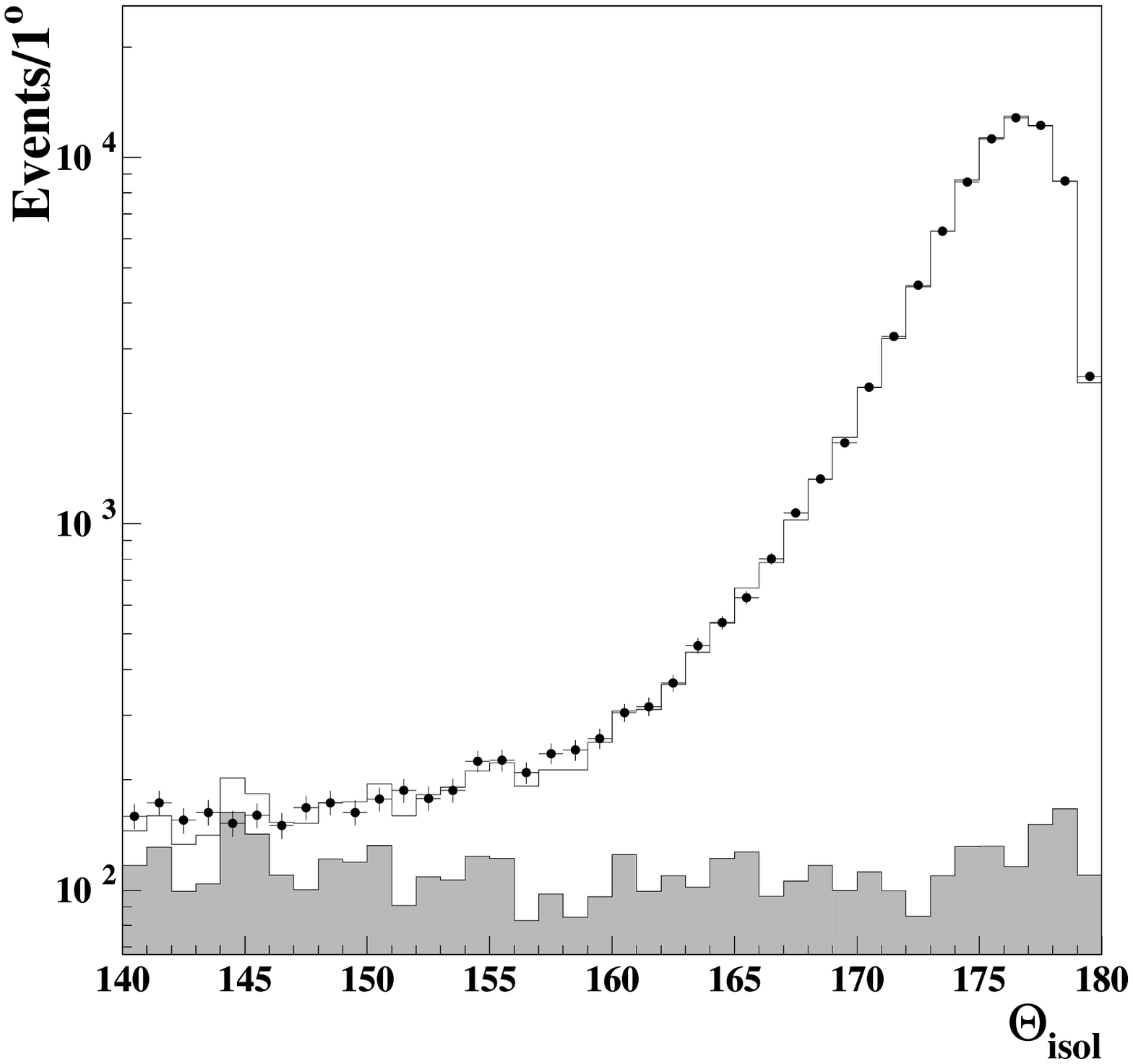,width=0.33\linewidth,height=7cm,bbllx=0,bblly=0,bburx=515,bbury=530}
\put(-120,160){\bf\huge c}    
    \caption{\it Distributions of isolation angle (in degrees) 
in candidate $\tt$ events: 
a) events containing two identified muons; 
b) events containing at least one tagged electron; 
c) all events in the $\tt$ sample. 
In all cases data are the dots and simulation is the  line. 
In (a) and (b) the shaded area is the four-fermion
background, while in (c) it contains all background contributions.} 
    \label{fig:isol}
  \end{center}
\end{figure}

The background due to the fully leptonic four-fermion final states was
estimated from the data  by extrapolating from the observed background
in  the low  isolation  angle region  for  events with  a low  visible
energy, as  well as for events tagged  by the presence of  a muon (for
$\ee\mm$)  or an  electron (for  $\ee\ee$).  The  relative uncertainty
obtained  on  the  normalisation  for  these final  states  was  $\pm$10\%.
Fig.~\ref{fig:isol}  shows the isolation  angle distributions  for 
different classes of events.

\begin{figure}[t]
  \begin{center}
  \vspace{-25pt}
\epsfig{figure=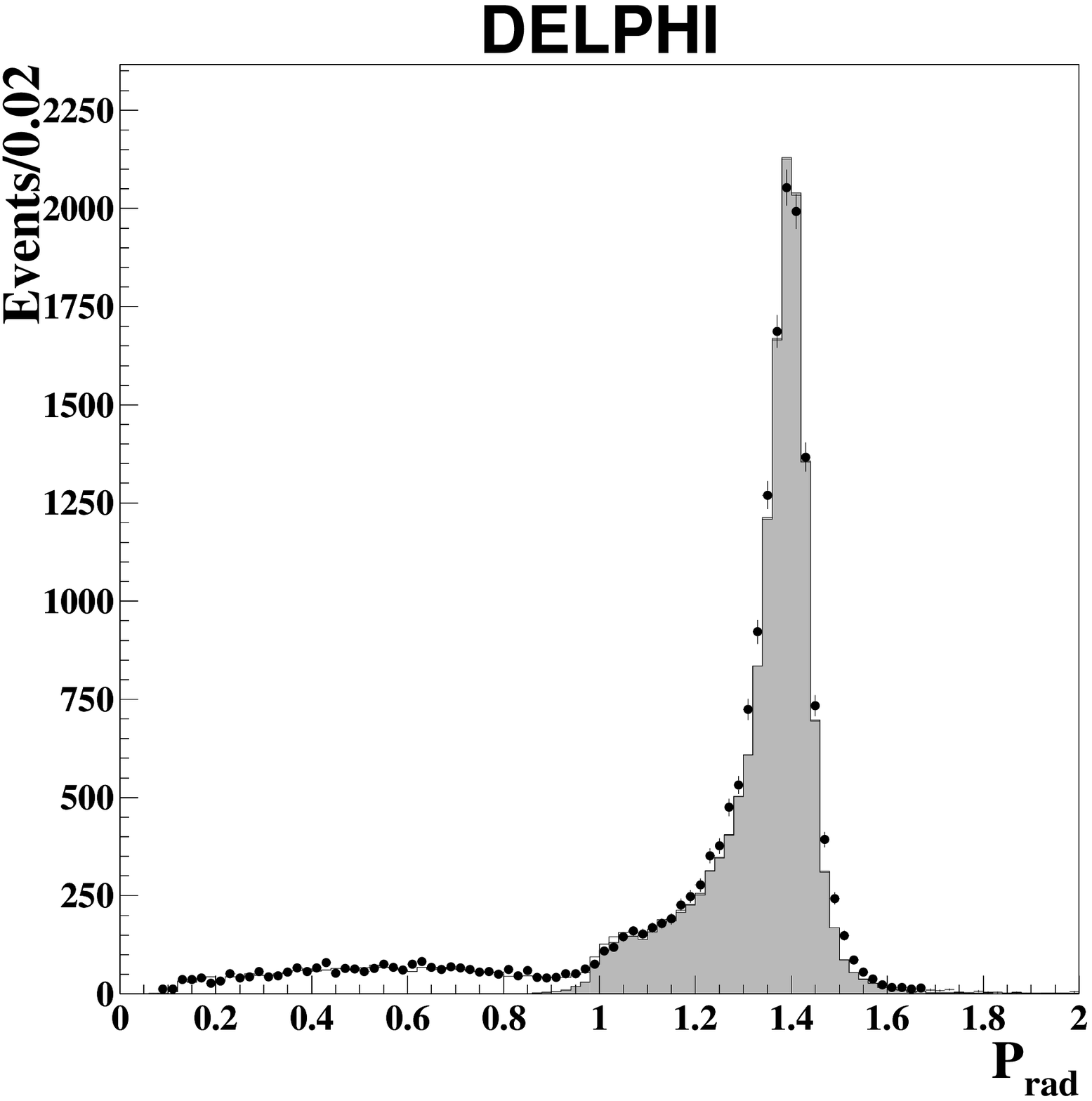,width=0.49\linewidth,height=8cm}
\put(-50,170){\bf\huge a}
\hfill
\epsfig{figure=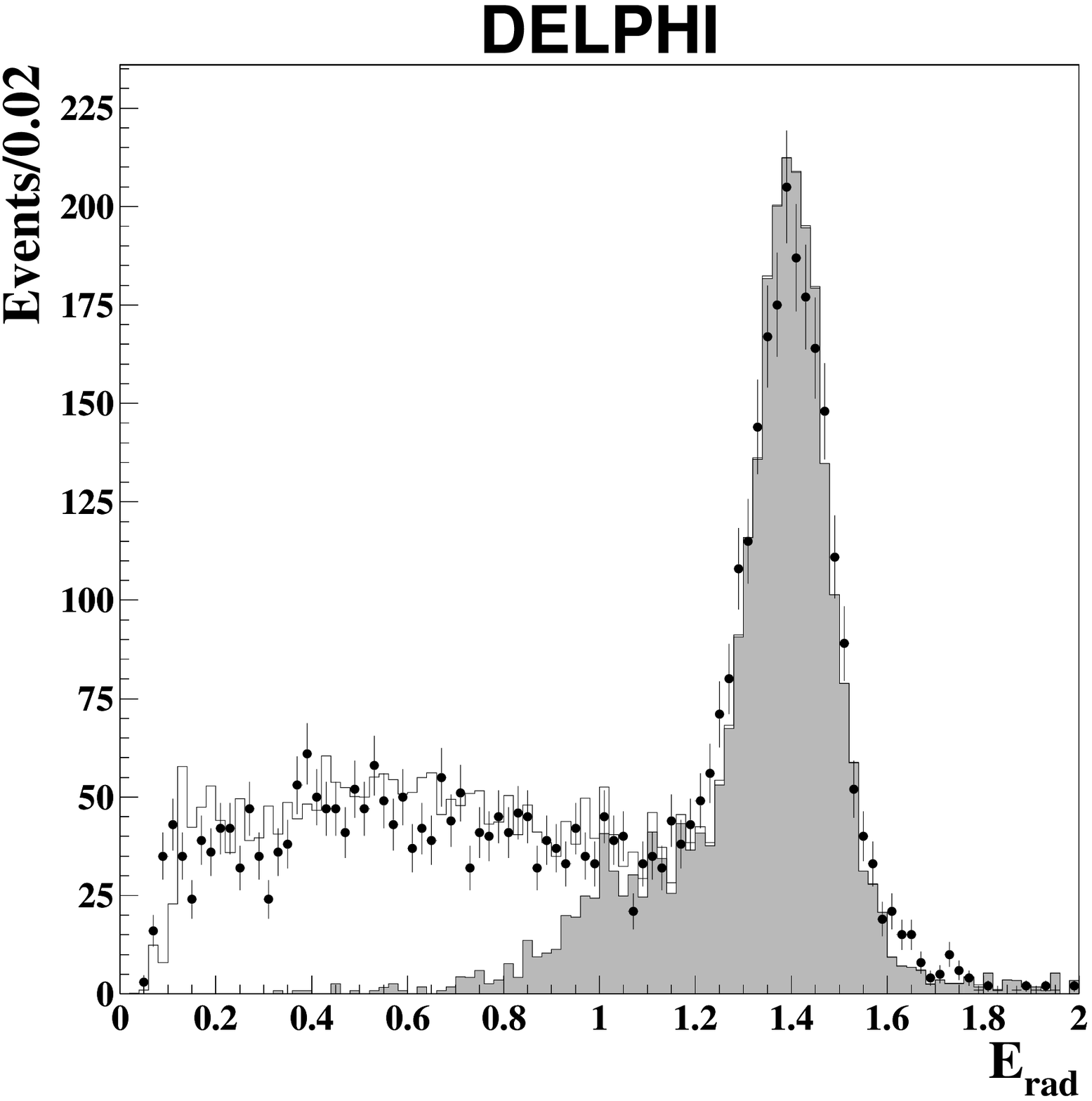,width=0.49\linewidth,height=8cm}
\put(-50,170){\bf\huge b}
    \caption{\it Distributions used in the study of the dileptonic
background to  the $\tt$  preselection: a) $p_{rad}$  distribution for
events containing two identified  muons; b) $E_{rad}$ distribution for
events containing  two identified electrons.   In both cases  the dots
are the data and the  line is simulation.  In (a) the shaded area
is the background contribution due to the reaction $\ee\to\mm(\gamma)$, 
while  in (b) it  is the
background contribution from $\ee\to\ee(\gamma)$ events.}
    \label{fig:radbkg}
  \end{center}
\end{figure}

The background from dimuon events in the $\tt$ sample was estimated by
studying the momentum distribution for high momentum identified muons,
and  by  studying the  $p_{rad}$  distributions  for  events with  two
identified muons.   This yielded a  correction factor relative  to the
simulation  of $0.96\pm0.03$.  The  $p_{rad}$ distribution,  after all
other  $\tt$ preselection  cuts, of  events containing  two identified
muons  is  shown   in~Fig.~\ref{fig:radbkg}a.   Similar  studies  were
carried out for Bhabha background using the associated electromagnetic
energy and  anti-electron tagging in  the hadron calorimeter  and muon
chambers.    This,  combined   with  studies~\cite{delphileptonicbr99}
comparing  the BHWIDE and  BABAMC Bhabha  generators indicated  that a
rescaling upwards by $1.15\pm0.15$  of the Bhabha background estimated
from simulation was necessary.  In addition, further cross-checks were
performed by studying the $p_{rad}$ and $E_{rad}$ distributions in the
region  outside  the one  occupied  by  $\tt$  events.  The  $E_{rad}$
distribution, after all other $\tt$ preselection cuts, for events with
two identified electrons is shown in~Fig.~\ref{fig:radbkg}b.

The  cosmic  ray contamination  was  estimated  from  the data  to  be
$(50\pm8)\times 10^{-5}$  by extrapolating  from the number  of events
seen outside the cuts in impact parameter.

The  total  background  was  estimated  to  be
$(1.51\pm0.10)\%$,  with  the  following breakdown:  $(0.11\pm0.01)\%$
from  $\mm$  final   states;  $(0.40\pm0.07)\%$  from  Bhabha  events;
$(0.29\pm0.01)\%$ from  $\qq$ events; $(0.27\pm0.03)\%$  from $\ee\ee$
final     states;    $(0.10\pm0.01)\%$    from     $\ee\mm$    events;
$(0.27\pm0.03)\%$ from  $\ee\tt$ final states;  $(0.02\pm0.01)\%$ from
$\ee\qq$  final  states;  $(0.05\pm0.01)\%$  from  cosmic  rays.   The
background  from  $\mm\mm$, $\mm\tt$  and  $\tt\tt$  final states  was
negligible.  

Satisfactory  agreement with the  simulation was observed
for each  year separately, and estimated  efficiencies and backgrounds
were compatible between the different years.

\boldmath
\section{Topology reconstruction} 
\unboldmath
\label{sec:reconstruction}
\boldmath
\subsection{Charged particle tracking} 
\unboldmath
\label{sec:tracking}
The charged particle  track pattern recognition was based  on a robust
algorithm designed to minimise  the number of bad associations between
hits produced by different particles  in the different elements of the
tracking  system.  In  general, most  tracks contain  hits in  all the
tracking subdetectors, the  VD, ID, TPC and OD.  Most importantly, for
the rejection of photon conversion products, the attachment of VD hits
to a track  must be as efficient and as  unambiguous as possible.  The
VD efficiency and misassociation  probability in simulation have been
tuned  to the  data  with $\eemm$  events  and with  large samples  of
hadronic    events     for    the    purposes     of    heavy    quark
tagging~\cite{btagging}.

The  TPC  was the  most  important  subdetector  for charged  particle
reconstruction.  The section  of a track reconstructed in  the TPC had
to contain at  least three space points out of  a maximum possible 16.
In  dense  topologies,  where  tracks  lay closer  together  than  the
inherent  two-track   resolution  of  the   detector,  the  clustering
algorithm  used  to  reconstruct  space points  used  the  information
contained in  the $\phi$  profile of a  single cluster to  resolve two
tracks down  to distances of  typically~2~to 3~mm in  r-$\phi$.  Where
tracks  were   closer  than  this,   a  space  point   was  associated
simultaneously to each of the  tracks with which it was compatible and
assigned a large error so as not to introduce systematic biases in the
reconstructed parameters for these  tracks.  For regions away from the
TPC sector  boundaries, the probability  to include either the  TPC or
the OD on an isolated track was close to 100\%.

After  the data  had been  passed  through the  standard algorithm,  a
second algorithm was run which produced VD-ID tracks from combinations
of hits in the VD and  the ID jet chamber.  This algorithm recuperated
charged  particle tracks  around the  boundary regions  of the  TPC or
cases  where the  two-track resolution  of the  TPC prevented  it from
resolving both  charged particles. The two-track resolution  of the VD
is more than  one order of magnitude better in  r-$\phi$ than the TPC.
However the VD-ID tracks had poorly reconstructed momenta and particle
charge  sign determination  due to  their  short length.  This had  no
consequences for the charged particle multiplicity reconstruction.

\boldmath
\subsection{Photon conversions} 
\unboldmath
\label{sec:conversions}
A good understanding of the  photons converting in material before the
tracking subdetectors  is important in order to  measure correctly the
true charged  particle multiplicity in  a $\tau$ decay.  About  7\% of
photons interact with  the material before the TPC  gas volume, giving
an  $\ee$  pair  detected  in   some  of  the  tracking  chambers.  In
particular,  unreconstructed conversions  form an  important potential
background from  lower multiplicity topologies  in higher multiplicity
samples.   Furthermore,  to  measure   in  data  both  the  conversion
reconstruction efficiency and the amount of material, it was necessary
to study the rate  of both reconstructed and unreconstructed converted
photons.

Converted  photons were reconstructed  using the  algorithms described
below.   These are  described in  more  detail in~\cite{delphipizero}.
Photons converting  in the beam-pipe, VD,  ID and TPC  inner wall were
reconstructed using charged particle  tracks observed in the TPC.  For
each TPC track  the position of the tangent to  the helix which passed
through the  interaction region was  calculated.  This point  gave the
estimated position of the conversion.  All tracks for which this point
was  compatible  with  the  interaction  region  within  one  standard
deviation were  neglected as conversion candidates.  If two conversion
candidates  were found  with compatible  conversion points,  they were
accepted  as a  converted photon  provided  that: the  two tracks  had
opposite curvature; the  mean conversion radius of the  two tracks was
greater than~5~cm; at least one  of the tracks had no associated point
inside the  mean conversion  radius; the angles of  the two
tracks  at the  conversion radii  agreed within  30 mrad  in $\phi$ 
and 15 mrad in $\theta$.
This algorithm had an  efficiency, estimated from simulation, of about
60\%  for  photon  energies  below  6~GeV falling  to  about  20\%  at
10~GeV. It  was complemented by  an algorithm using  a fit to  a joint
origin for  two charged particle  tracks assuming zero  opening angle,
with  similar  requirements  on  the  position  of  the  reconstructed
conversion radius to the previous algorithm. This had an efficiency of
about 40\%  for photon energies  above 8~GeV.  The  average efficiency
for the  combination of both  algorithms was $(68.1\pm0.2)\%$  for the
photon  energy  spectrum  in   simulated  1-prong  $\tau$  decays  and
$(59.8\pm0.4)\%$ for 3-prong decays.

In 3-prong  decays, the  probability of reconstructing  a $\pi^+\pi^-$
pair incorrectly as a  photon conversion was estimated from simulation
to  be $1\times10^{-4}$.   For systematic  studies the  uncertainty on
this was taken as 100\% of the size of the effect.  The probability of
reconstructing  a $\kos\to\pi^+\pi^-$  decay incorrectly  as  a photon
conversion was estimated from simulation to be 1.5\%, however this did
not alter the observed charged particle multiplicity.

The rates of reconstructed  and unreconstructed converted photons were
studied  as a  function  of  the radial  coordinate  of the  estimated
conversion  point. The  conversions were  separated into  those taking
place in the  beam-pipe, VD and ID jet chamber inner  wall, the ID jet
chamber sensitive volume, the ID jet chamber outer wall and ID Trigger
Layers, and  the TPC inner wall. The  bulk of the material  was in the
TPC inner  wall.  In  order to  be as independent  as possible  of the
$\tt$ presample and modelling of  the photon production rate in $\tau$
decays, the study was  performed using $\eemm\gamma$ and $\bhab\gamma$
radiative  events.  The events  were selected  using electron  or muon
identification and used kinematic constraints to predict the direction
and energy of the photon.  The  rates were recorded for events with an
HPC shower,  reconstructed conversion, or  no reconstructed conversion
but some charged particle  tracks consistent with an electron-positron
pair arising from a converted photon. In the latter case the radius of
the conversion  point was  taken as being  that of the  material layer
lying just inside the innermost  first measured point on either of the
two  tracks believed  to be  due to  the conversion.   The  rates were
normalised  to  the  total   number  of  observed  photons,  including
conversions.   The   identical    procedure   was   applied   to   the
simulation. Correction  factors for the  reconstructed conversion rate
were  obtained  from the  ratio  $A^{^{conv}}_{_{rec}}$  of the  rates
$R^{^{rec}}_{d}$ and $R^{^{rec}}_{s}$  of reconstructed conversions in
data and simulation respectively. Analogous correction factors for the
unreconstructed   conversion  rate  were   obtained  from   the  ratio
$A^{^{conv}}_{_{unrec}}$   of   the   rates   $R^{^{unrec}}_{d}$   and
$R^{^{unrec}}_{s}$ of unreconstructed conversions.

The   measured  correction   factors   were  used   to  estimate   the
reconstruction   efficiency  and  amount   of  material.    The  ratio
$A^{^{conv}}_{_{rec}}$ is given~by
\begin{equation}
A^{^{conv}}_{_{rec}} = \frac{R^{^{rec}}_d}{R^{^{rec}}_s} =
\frac{R^0_d}{R^0_s} \cdot \frac{X^{^{EM}}_d}{X^{^{EM}}_s} 
 \cdot \frac{\epsilon^{^{conv}}_d}{\epsilon^{^{conv}}_s}.
\label{egn:conv1}
\end{equation}
The subscripts $d$ and $s$ stand for data and simulation respectively.
$R^0_{d,s}$  is the total  rate of  photons, $X^{^{EM}}_{d,s}$  is the
probability of  a conversion taking place (proportional  to the amount
of     material    in    terms     of    radiation     lengths)    and
$\epsilon^{^{conv}}_{d,s}$  is  the  reconstruction  efficiency.   The
ratio $A^{^{conv}}_{_{unrec}}$ is given~by
\begin{equation}
A^{^{conv}}_{_{unrec}} = \frac{R^{^{unrec}}_d}{R^{^{unrec}}_s} =
\frac{R^0_d}{R^0_s} \cdot \frac{X^{^{EM}}_d}{X^{^{EM}}_s} 
 \cdot \frac{1-\epsilon^{^{conv}}_d}{1-\epsilon^{^{conv}}_s}.
\label{egn:conv2}
\end{equation}
The conversion reconstruction efficiency in data $\epsilon^{^{conv}}_d$ can
be obtained from the data alone via the relation
\begin{equation}
\epsilon^{^{conv}}_d = \frac{R^{^{rec}}_d}{R^{^{rec}}_d+R^{^{unrec}}_d},
\label{egn:conv3}
\end{equation}
independently of any assumptions on the material or, if $R^{^{rec}}_d$
and  $R^{^{unrec}}_d$ are  measured on  the same  data sample,  on the
incident  photon   flux.   If   the  ratio  $R^0_d/R^0_s$   is  known,
measurements  of $A^{^{conv}}_{_{rec}}$,  $A^{^{conv}}_{_{unrec}}$ and
$\epsilon^{^{conv}}_d$, combined  with $\epsilon^{^{conv}}_s$ from the
simulation,  can   provide  an   estimation  of  the   material  ratio
$X^{^{EM}}_d/X^{^{EM}}_s$ for electromagnetic interactions.  The ratio
$R^0_d/R^0_s$ of  the rates of  photons produced in  the $\eemm\gamma$
and  $\bhab\gamma$ radiative  events  was  taken to  be  unity with  a
systematic uncertainty of~$\pm$2\%~\cite{bjarnemugam}.

\begin{figure}[t]
  \begin{center}
  \vspace{-25pt}
\epsfig{figure=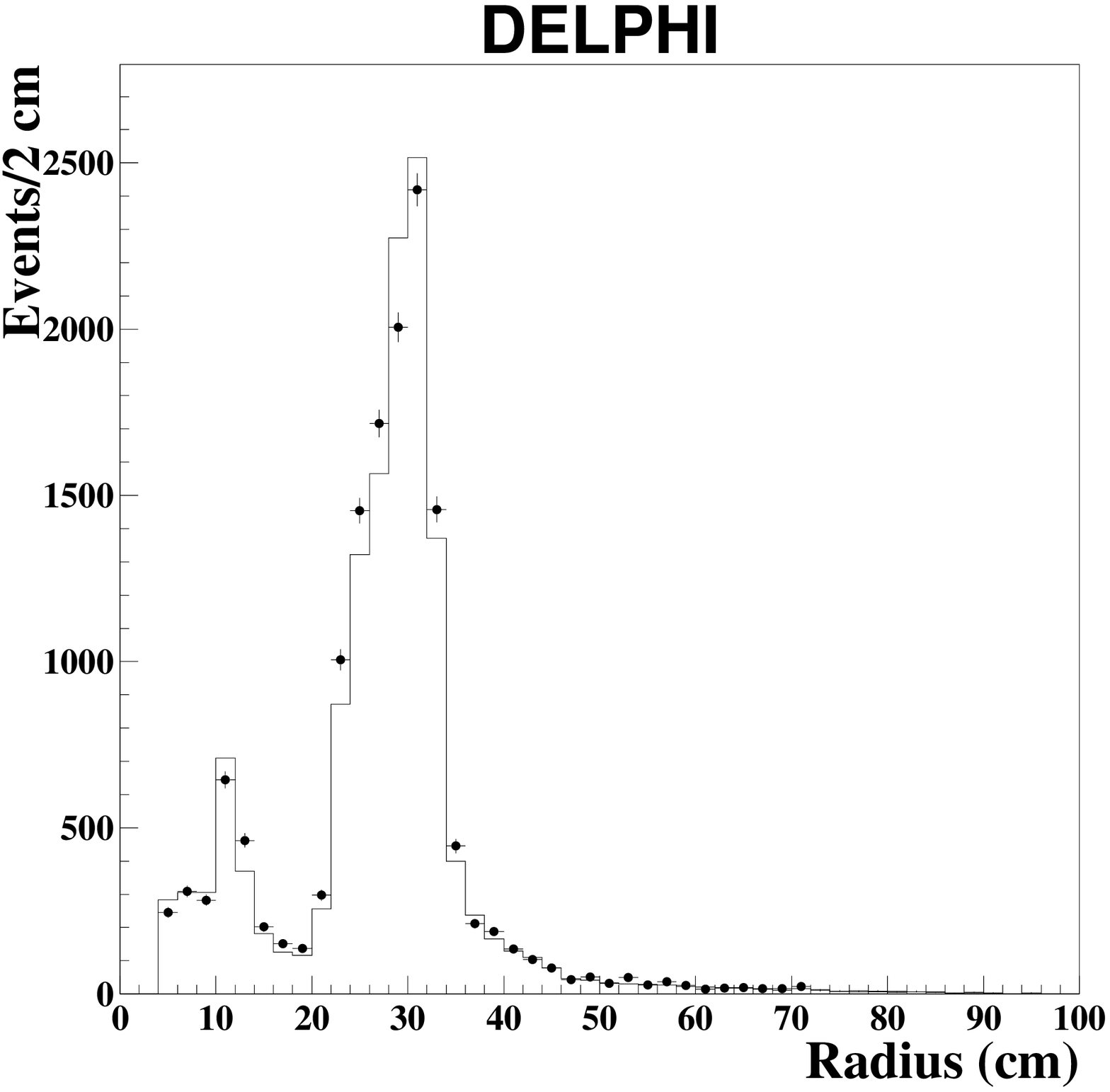,width=0.49\linewidth,height=8cm}
\put(-50,170){\bf\huge a}
\hfill
\epsfig{figure=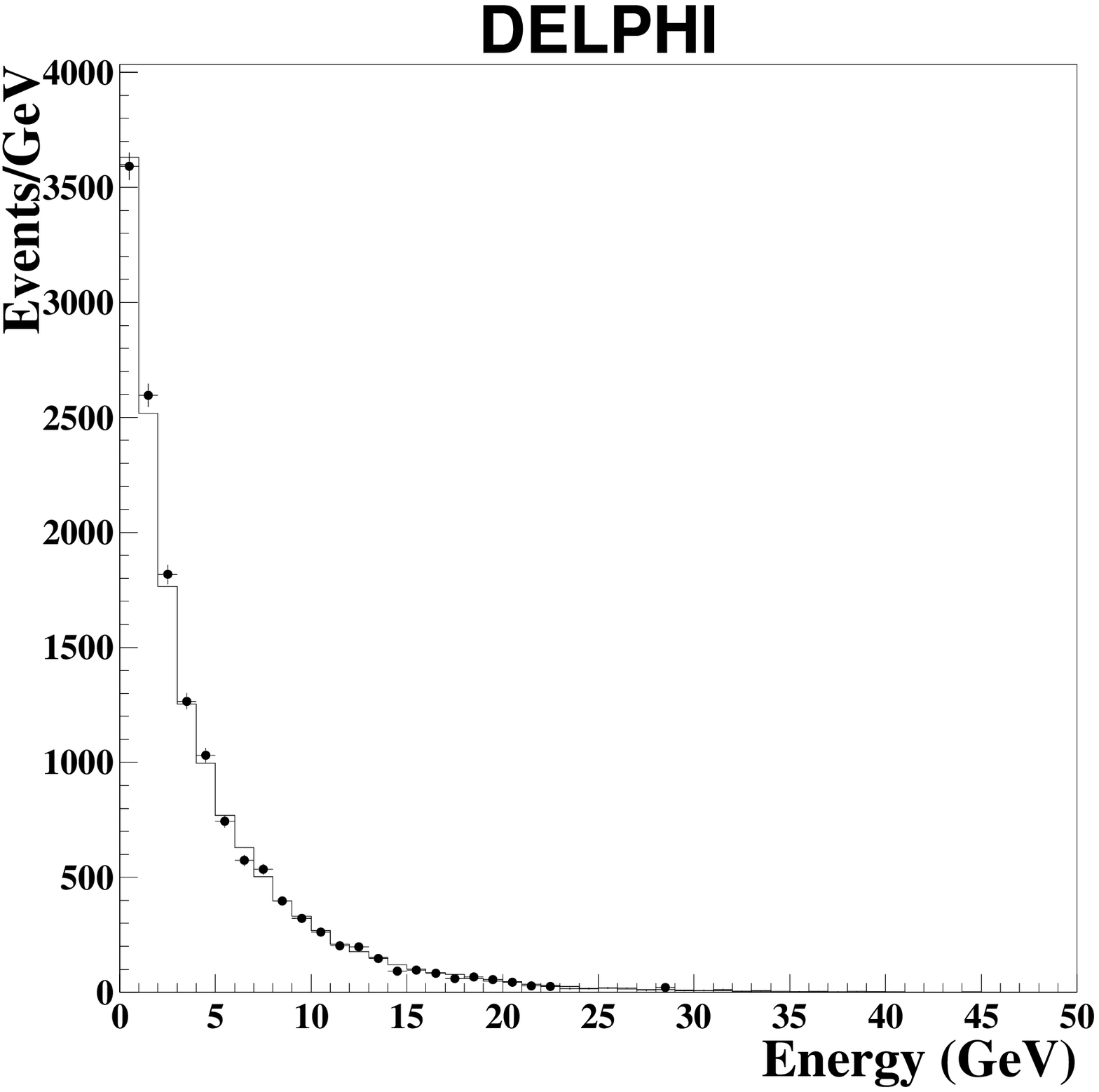,width=0.49\linewidth,height=8cm}
\put(-50,170){\bf\huge b}
        \caption{\it a) Distribution in $\eett$ events of the 
             estimated  conversion  radius  for photons  converted  in
             front  of the  TPC.  Data  are dots,  the   line  is
             simulation   after  applying   correction   factors  from
             $\ee\to\mm\gamma$     and    $\ee\to\ee\gamma$    events.
             b)~Reconstructed  conversion  photon  energy  in  $\eett$
             events.   Data are  dots,  the   line is  simulation
             after applying correction~factors.}
    \label{fig:conv1}
  \end{center}
\end{figure}

The   factors   $A^{^{conv}}_{_{rec}}$  and   $A^{^{conv}}_{_{unrec}}$
obtained   as  a  function   of  interaction   radius  are   given  in
Table~\ref{tab:conversion_corrections}.  These  factors  were used  to
reweight the simulated events used  in the estimation of the selection
efficiencies  to  account  for  data/simulation discrepancies  in  the
material  and  reconstruction efficiency.   The  distributions of  the
conversion radius  and energy for reconstructed  converted photons are
shown in Fig.~\ref{fig:conv1}, where the simulation has been corrected
with  the  factors obtained  in  the  $\eemm\gamma$ and  $\bhab\gamma$
events.  There are still some localised discrepancies between data and
simulation due to simplifications in the detector material description
in  the simulation  program. For  the  $\tau$ presample,  the rate  of
reconstructed conversions  observed in data  was 1.004$\pm$0.008 times
that  predicted  by  the  simulation  after  the  application  of  the
correction factors, in excellent agreement.
\begin{table}[t]
\renewcommand{\tabcolsep}{0.5pc} 
\renewcommand{\arraystretch}{1.2} 
\begin{center}
\vspace{-10pt}
\begin{tabular}{l|l|r@{$\pm$}l|r@{$\pm$}l|r@{$\pm$}l|r@{$\pm$}l}
\hline 
\multicolumn{2}{c}{Radius} &\multicolumn{8}{|c}{Conversions} \\
 \cline{1-10}   
  \multicolumn{1}{c}{Lower}  &\multicolumn{1}{|c}{Upper}
 &\multicolumn{2}{|c}{$A^{^{conv}}_{_{rec}}$}
 &\multicolumn{2}{|c}{$A^{^{conv}}_{_{unrec}}$}
 &\multicolumn{2}{|c}{$\epsilon^{^{conv}}_{d}/\epsilon^{^{conv}}_{s}$}
 &\multicolumn{2}{|c}{$X^{^{EM}}_{d}/X^{^{EM}}_{s}$}  \\ 
\hline  
   0cm & 13.0cm & 1.00& 0.03 & 1.14& 0.05 & 0.96& 0.03 & 1.04& 0.03 \\
 13.0cm& 22.0cm & 1.34& 0.05 & 1.70& 0.12 & 0.93& 0.06 & 1.45& 0.05 \\
 22.0cm& 28.5cm & 1.23& 0.03 & 1.50& 0.09 & 0.94& 0.05 & 1.31& 0.03 \\
 28.5cm& 50cm   & 1.04& 0.02 & 0.91& 0.04 & 1.04& 0.03 & 1.00& 0.02 \\ 
\hline 
 0cm   &  50cm  & 1.10& 0.01 & 1.09& 0.03 & 1.00& 0.02 & 1.09& 0.02 \\
 \hline
\end{tabular}
\end{center}
\caption{\it  Factors used to correct simulated photon conversions 
and data/simulation discrepancies  as a function of the  radius of the
interaction  point.   Column  5  shows the  data/simulation  ratio  of
reconstruction efficiency  and column 6  the equivalent ratio  for the
material.}
\label{tab:conversion_corrections}
\end{table}
The data taken in different years were studied separately, in particular
before and after the beginning of 1995, the point after which
the material in the ID trigger layers was substantially reduced.
Consistency was observed between the different years and the
change in conversion rate observed in the ID trigger layers
between 1992-1994 and 1995 was consistent with the expectation.

The  conclusion, integrating  up to  a radius  of 50~cm,  is  that the
material is underestimated in the detector simulation by about 10\% in
terms  of  radiation  lengths,  while  the  conversion  reconstruction
algorithm efficiency is similar in data and simulation.

The   kinematical  conversion   reconstruction  was   complemented  by
conversion   rejection    based   on   an    electron   identification
algorithm~\cite{delphi_performance}  which  combined information  from
the TPC ionisation and  the HPC electromagnetic shower reconstruction.
This  algorithm also rejected  many $\pi^0$  Dalitz decays,  which was
particularly important for rejecting backgrounds in the 5-prong $\tau$
decay  sample.  A conversion  was flagged  if there  was at  least one
tightly tagged electron in a hemisphere with greater than one track in
the TPC  or if there  were a pair  of oppositely charged particles  both of
which were loosely tagged as electrons.  These criteria were chosen to
minimise  the probability  of  misidentifying a  $\pi^+\pi^-$ pair  as
$\ee$   rather  than   to  maximise   the   conversion  identification
efficiency.  In simulation,  the  efficiency to  reject the  remaining
conversions    was   (59.6$\pm$0.6)\%    in    1-prong   decays    and
(54.5$\pm$0.8)\%  in  3-prong   decays.   The  electron-positron  pair
rejection also identified (72.7$\pm$1.2)\% of Dalitz pairs from $\pio$
decays in simulated 1-prong  decays, giving an improved classification
for  these decays.   The efficiency  was (68.3$\pm$3.1)\%  for 3-prong
decays.  The efficiency  to identify Dalitz pairs was  higher than for
conversions  due  to  biases  arising from  the  kinematic  conversion
reconstruction   algorithm   which    had   already   identified   and
reconstructed most converted photons.

The  conversion rejection  efficiency of  the  electron identification
algorithm was cross-checked in  $\tau$ decays containing three charged
particle  tracks, from the  rate of oppositely charged pairs  of particles
which satisfied the  electron tagging requirements outlined  above and where
neither particle had a hit in  the VD.  The ratio of the efficiency in
data to that in simulation  was estimated to be 0.96$\pm$0.05, in good
agreement  with  unity. The  relative  systematic  uncertainty on  the
efficiency of this  algorithm was conservatively taken to  be 10\%, to
account for any systematic effects  in the estimation of the ratio and
in its application to all classes of charged particle tracks.

\begin{figure}[t]
  \begin{center}
  \vspace{-25pt}
\epsfig{figure=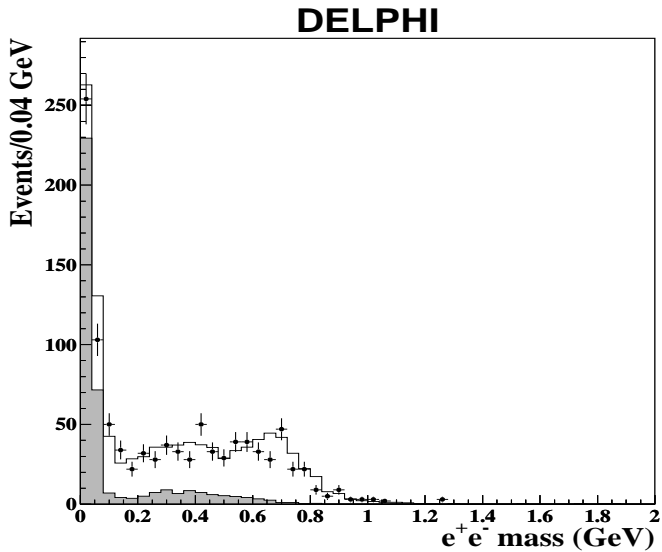,width=0.6\linewidth,height=8cm}
\caption{\it Invariant mass $m_{\ee}$ distributions 
for pairs of oppositely  charged  particles   with  one  or  both  particles
identified as electron  candidates and where both tracks  have hits in
the~VD. Data  are dots, the simulation  is the   line. The shaded
area  shows  the  contribution  from $\pi^0\to\gamma\ee$  decays.  The
region      $m_{\ee}>{\mathit{200}}$~MeV      is     dominated      by
misidentified~$\pi^+\pi^-$~pairs.}
    \label{fig:dalitz1}
  \end{center}
\end{figure}
Fig.~\ref{fig:dalitz1} shows the invariant mass distribution for pairs
of oppositely  charged   particles  in  $\tau$  decay  hemispheres
containing three charged particles, where one or both of the pair were
tagged as an electron candidate and  where both tracks had hits in the
VD.   The  electron mass  was  assumed  for  both particles.   The  VD
association  requirement  rejected   most  photon  conversions,  which
occurred  after the  VD, while  not  affecting the  signal of  $\pi^0$
Dalitz  decays,  or  of   hadrons  misidentified  as  electrons.   The
contribution from Dalitz decays peaks at low masses, below the $\pi^0$
mass.   The tail from  Dalitz decays  in the  region of  masses around
200~MeV to 600~MeV  arises from decays where only  one of the electron
or positron was identified and there were two possible combinations of
oppositely charged particles. In such decays only half the possible
combinations  were  due  to  the  $\ee$  pair.  In  these  cases  each
combination has been  given a weight of one half.   The region of mass
greater than 200~MeV was dominated  by $\pi^+\pi^-$ pairs where one of
the  pions  was  misidentified  as  an  electron.   This  permitted  a
determination  from  the data  of  the  probability  to misidentify  a
3-prong  $\tau$  decay as  a  1-prong  decay  containing a  conversion
candidate.   The  misidentification probability  was  estimated to  be
$(1.29\pm0.09)\%$.  The uncertainty was  dominated by  data statistics
and contained a contribution from the VD association efficiency.  From
the ratio of data to simulation in the low mass spike it was estimated
that  the efficiency  to reject  a $\pi^0$  Dalitz decay  in  data was
0.90$\pm$0.07  times  the   efficiency  obtained  in  simulation.  The
uncertainty included contributions from the error on the Particle Data
Group   best   estimation   of   the   $\pi^0\to\gamma\ee$   branching
ratio~\cite{pdg2000},  the uncertainty  in the  VD efficiency  and the
limited   simulation  statistics  in   addition  to   the  statistical
uncertainty.  No  correction  was  applied  to the  simulation  and  a
relative uncertainty  of 10\% on  the Dalitz rejection  efficiency was
taken.

Most $\delta$-rays  were rejected by the  requirement of VD  hits on a
track. Events in simulation  containing a $\delta$-ray were reweighted
with the radius dependent factors  describing the ratio of material in
terms      of      radiation     lengths~$X^{^{EM}}_{d}/X^{^{EM}}_{s}$
in~Table~\ref{tab:conversion_corrections}. There is some ambiguity
in the choice of this factor. The associated systematic uncertainties
are discussed in Section~\ref{sec:systematics_reinteractions}.

\boldmath
\subsection{Nuclear reinteractions} 
\unboldmath
\label{sec:nucints}
About 3\% of hadrons reinteract inelastically with the material before
the TPC gas  volume, and produce typically up  to 10 charged secondary
particles.   These reinteractions were  reconstructed by  an algorithm
which was designed to  find secondary reinteraction vertices using the
tracks   from   outgoing  charged   particles   produced  in   nuclear
reinteractions.

The  algorithm  was robust  with  respect  to  the number  of  charged
particles produced and  to the position of the  reinteraction with the
DELPHI  detector material.   Only tracks  which had  impact parameters
inconsistent with coming from  the interaction region were considered.
The algorithm iterated over all  pairs of tracks, and attempted to fit
a vertex in three dimensions if neither of the tracks had already been
included in an earlier successfully fitted vertex.  However, if one of
the tracks  had already been included  in an earlier  fitted vertex, an
attempt was made to include the  other track in that vertex.  This was
the  main  part of  the  algorithm  permitting  the reconstruction  of
vertices with an arbitrary number of outgoing charged particle tracks.
In  the rare case  where both  tracks had  already been  associated to
different  fitted vertices,  an attempt  was made  to merge  those two
vertices,  if  their  positions  were  compatible.   In  all  cases  a
candidate  reinteraction vertex  had to  pass a  $\chi^2$ cut  and all
outgoing tracks had  to have a first observed  hit consistent with the
particle  originating   from  the  reconstructed   vertex.   Candidate
vertices with  two outgoing  tracks and kinematically  compatible with
the decay of a $\kos$  or $\Lambda$ produced in the interaction region
were not considered.

A search was then performed for the incoming particle which caused the
interaction.  Tracks  which had been  reconstructed in only the  VD or
the VD and ID were extrapolated to the candidate nuclear reinteraction
vertex, and  associated to  the vertex if  consistency was  found. 
This prevented  double counting of VD-ID tracks
and reconstructed nuclear reinteractions. If
there was no such track the momentum and charge of the incoming hadron
were estimated from the sum  over the outgoing particles associated to
the vertex.   For $\tau$ decays it  is a good  approximation to assume
all  such  nuclear  reinteraction   vertices  are  caused  by  charged
particles,  as $K^0$  mesons are  the only  source of  neutral hadrons
producing nuclear reinteractions and are infrequently produced.

In simulation, the algorithm was (82.4$\pm$0.2)\% efficient in 1-prong
$\tau$  decays  for incoming  particle  momenta  above 1~GeV/$c$,  and
(83.2$\pm$0.4)\% efficient  in 3-prong $\tau$  decays.  The efficiency
had  little dependence  on the  $\tau$  decay multiplicity  or on  the
charged secondary particle  multiplicity in the nuclear reinteraction.
The position of  the reinteraction  vertex was reconstructed
with a precision of better than 1~mm in three dimensions.
\begin{figure}[t]
  \begin{center}
  \vspace{-25pt}
\epsfig{figure=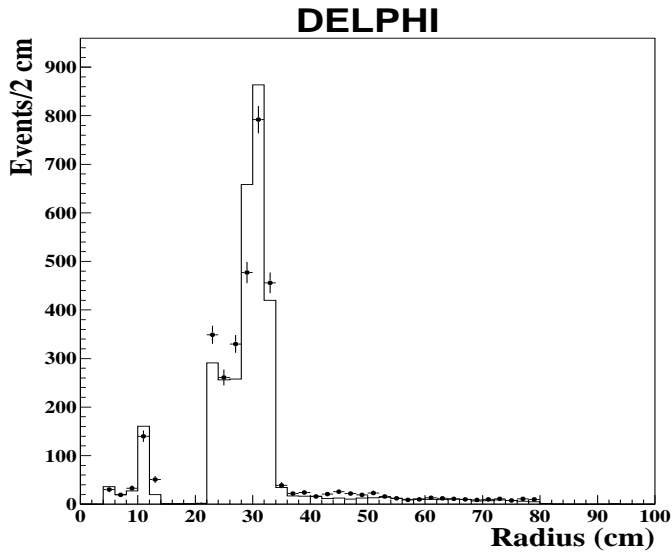,width=0.6\linewidth,height=8cm}
\caption{\it Distribution in $\eett$ events of reconstructed radius
             of the inelastic  nuclear reinteractions.  Data are dots,
             the   line is  simulation after  applying correction
             factors.  }
    \label{fig:hadint1}
  \end{center}
\end{figure}

In a  similar manner as  for the converted photons,  the reconstructed
and unreconstructed nuclear reinteractions  were studied as a function
of  the  reinteraction radius.   The  rates  of reconstructed  nuclear
reinteractions  were  obtained  from  the  $\tt$ sample  in  data  and
simulation.   The   radial  distribution  of   reconstructed  hadronic
reinteractions  is shown  in  Fig.~\ref{fig:hadint1}.  Some  localised
discrepancies, more marked than in the case of the photon conversions,
are visible in  the detailed description of the  ID trigger layers and
TPC in the simulation. The flat tail
for radii greater than 35~cm is due to reinteractions in the TPC
cathode plane at~$\theta\approx 90^\circ$.

The unreconstructed  nuclear reinteractions were  measured by studying
the multiplicity distribution for  charged particle tracks with impact
parameters in the r-$\phi$ plane greater than 1.5~cm in the hemisphere
opposite    a    well     reconstructed    leptonic    $\tau$    decay
candidate. Simulation studies showed that this was dominated by $\tau$
decays   with    an   unreconstructed   nuclear    reinteraction   for
multiplicities of four and greater.  The $\tau$ decay hemispheres used
to  estimate  this  effect   had  to  have  no  reconstructed  nuclear
reinteraction.  This  removed the risk of a  false measurement arising
from potential discrepancies in  the charged particle multiplicity due
to data/simulation  differences in  the track inclusion  efficiency of
the nuclear reinteraction reconstruction algorithm.  The reinteraction
radius  was  taken  as  the  radius  of  the  material  surface  lying
immediately  inside the  first measured  point on  any of  the charged
particle tracks with a high impact parameter.

\begin{table}[t]
\renewcommand{\tabcolsep}{0.5pc} 
\renewcommand{\arraystretch}{1.2} 
\begin{center}
  \vspace{-10pt}
\begin{tabular}{l|l|r@{$\pm$}l|r@{$\pm$}l|r@{$\pm$}l|r@{$\pm$}l}
\hline    \multicolumn{2}{c}{Radius}    &\multicolumn{8}{|c}{Inelastic
 nuclear  reinteractions}  \\  \cline{1-10}  \multicolumn{1}{c}{Lower}
 &\multicolumn{1}{|c}{Upper}
 &\multicolumn{2}{|c}{$A^{^{nucl}}_{_{rec}}$}
 &\multicolumn{2}{|c}{$A^{^{nucl}}_{_{unrec}}$}
 &\multicolumn{2}{|c}{$\epsilon^{^{nucl}}_{d}/\epsilon^{^{nucl}}_{s}$}
 &\multicolumn{2}{|c}{$X^{^{nucl}}_{d}/X^{^{nucl}}_{s}$} \\ \hline 0cm
 & 22.0cm  & 0.91& 0.08 &  1.01& 0.02 & 0.95  & 0.05 & 0.96  & 0.04 \\
 22.0cm& 28.5cm & 0.93& 0.04 & 1.36&  0.15 & 0.96 & 0.13 & 0.97 & 0.04
 \\ 28.5cm& 50cm & 0.86& 0.03 & 1.00& 0.03 & 0.98 & 0.03 & 0.87 & 0.03
 \\ \hline 0cm & 50cm & 0.89& 0.02 & 1.01& 0.02 & 0.98 & 0.02 & 0.91 &
 0.02 \\ \hline
\end{tabular}
\end{center}
\caption{\it Factors used to correct simulated nuclear reinteractions
for data/simulation discrepancies  as a function of the  radius of the
reinteraction  point.  Column  5  shows the  data/simulation ratio  of
reconstruction efficiency  and column 6  the equivalent ratio  for the
material.}
\label{tab:reinteraction_corrections}
\end{table}
The    measured   correction   factors    $A^{^{nucl}}_{_{rec}}$   and
$A^{^{nucl}}_{_{unrec}}$,     analogous      to     the     quantities
$A^{^{conv}}_{_{rec}}$  and $A^{^{conv}}_{_{unrec}}$  for conversions,
are given  in~Table~\ref{tab:reinteraction_corrections}, together with
the data/simulation  ratio of  the efficiencies for  reconstruction of
inelastic                                                       nuclear
reinteractions~$\epsilon^{^{nucl}}_{d}/\epsilon^{^{nucl}}_{s}$ and the
derived  data/simulation  ratio  of   material  in  terms  of  nuclear
interaction lengths~$X^{^{nucl}}_{d}/X^{^{nucl}}_{s}$.   The region of
the  ID jet  chamber gas  volume, (13~cm  to~22~cm) has  been included
together with the region at lower radius as it has a very small number
of  nuclear interaction  lengths, of  order~$10^{-4}$.   These results
indicate  that for nuclear  interactions the  material at  lower radii
than  the  TPC  gas volume  is  over-estimated  by  about 9\%  in  the
simulation program  while the reconstruction  efficiency is consistent
in  data and  simulation.  There  is a  significant difference  in the
estimated   data/simulation   material    ratios   for   nuclear   and
electromagnetic   interactions.   This  can   be  attributed   to  the
complexity of the composite material structure together with the means
by  which each  of these  quantities is  introduced in  the simulation
program, as  well as  to possible weaknesses  in the modelling  of the
hadronic reinteractions  in the simulation program.   The same effects
are  observed  in $\eeha$  events.   The  uncertainty  in the  initial
production rate  of hadrons in  $\tau$ decays is relatively  small and
does not  give an important systematic uncertainty  in the measurement
of $A^{^{nucl}}_{_{rec}}$ and $A^{^{nucl}}_{_{unrec}}$.

As for the photon conversions, the nuclear reinteractions
were studied for each year separately, particularly before and after
the upgrade of the ID trigger layer in 1995.

Simulation studies  indicate that the probability  of reconstructing a
fake nuclear reinteraction from $\tau$ decay secondary particles where
none  was present  was $5  \times 10^{-4}$  in 3-prong  $\tau$ decays.
However  most such  events  were still  classified  as 3-prongs.   The
associated  uncertainty was  taken  to be  100\%  of the  size of  the
effect.    The   probability   of   reconstructing  a   fake   nuclear
reinteraction from a $\kos$ decay was negligible.
%

The probability that a hadron produced in a $\tau$ decay would undergo
an elastic  nuclear reinteraction before the TPC  sensitive volume was
estimated  from  simulation  to  be  (3.29$\pm$0.05)\%.   The  typical
signature of such  an interaction was a charged  particle track with a
kink  at the  scattering point  in the  detector material.   For large
scatters the reconstruction algorithm would not reconstuct one but two
tracks, one before and one after  the scatter. This led typically to a
signature of one  VD-ID track and one TPC-OD track,  and did not alter
the observed track multiplicity in the TPC and OD.

\boldmath
\subsection{Topology selection  criteria}
\label{sec:topology}
\unboldmath By  judicious application of the VD  hits requirement, and
taking into account  that the $\tau$ must decay into  an odd number of
charged  particles,  algorithms   were  constructed  to  minimise  the
sensitivity to the VD association  efficiency and the knowledge of the
material reinteractions; these are described below. In the following a
``good''  track is  defined  as  a charged  particle  track which  has
associated TPC or OD hits and  is not identified as originating from a
conversion  or nuclear  reinteraction  in the  detector material.  For
simplicity, the  VD-ID track classification, in  addition to including
tracks  reconstructed using  only the  VD and  ID  detectors, includes
tracks  reconstructed  as the  ingoing  charged  hadron  to a  nuclear
reinteraction.  In $\tau$ decay  hemispheres where  the presence  of a
converted  photon   was  signalled  by   the  electron  identification
algorithm, the observed  number of good tracks was  reduced by two (or
by one  in the rare  case of  an even number  of good tracks)  for the
purposes of the topology classification.

Individual $\tau$  decay hemispheres  were allotted to  five different
classes dependent  on the number  of reconstructed tracks in  the TPC,
VD,  etc.  The  first three  of  these classes  correspond closely  to
events  with clean  reconstruction of  a 1-prong,  3-prong  or 5-prong
topology.  The  other two classes  contain the small number  of $\tau$
decay hemispheres where there was more ambiguity in the reconstruction
of the topology, but where some discrimination was still possible.

A  1-prong $\tau$  decay  was  defined as  a  $\tau$ decay  hemisphere
satisfying any of the following criteria:
\begin{itemize}
\item 
only one good track  with associated VD hits, and no other 
tracks with associated VD hits;
\item
only one good track, without VD or ID hits, and one VD-ID track;
\item
no good tracks, and only one VD-ID track.
\end{itemize}
3-prong $\tau$ decays  were  isolated by  demanding $\tau$ decay 
hemispheres satisfying at least one of the following sets of criteria:
\begin{itemize}
\item  
three, four or five good tracks, of which either two or three 
had VD hits;
\item 
two good tracks with associated VD  hits, plus one VD-ID track;
\item
one good track with associated VD  hits, plus one or two
VD-ID tracks pointing within $3^\circ$ in azimuth of a TPC sector
boundary.
\end{itemize}
Candidate  5-prong $\tau$ decays   were selected  if they
satisfied at least one of the following topological criteria:
\begin{itemize}
\item 
five good tracks of which at least four had two or more 
associated VD hits;
\item 
four good tracks with VD hits, and one other VD-ID track.
\end{itemize}
Additional criteria  were applied in  the selection of  5-prong $\tau$
decays due to the large  potential background from hadronic $Z$ decays
and   mis-reconstructed   3-prong   $\tau$  decays.   The   background
originating from $\nut 3h^\pm \ge  1 \pi^0$ final states with a Dalitz
decay was  expected to occur  at a similar  level to the  signal.  The
electron rejection criteria described in Section~\ref{sec:conversions}
reduced this background  by about 70\%, and it  was further suppressed
by   requiring  that   all  good   tracks  had   a   momentum  greater
than~1~GeV/$c$.  To  reject $Z\to\qq$ events it was  required that the
total   momentum  of   the   5-prong  system   be   greater  than
20~GeV/$c$. Only good tracks were  included in the calculation of this
quantity.

These three classes accounted for 97.6\% of candidate $\tau$ decays in
the $\tt$ sample.  The remaining 2.4\% of candidate $\tau$ decays were
mostly 1-prong and 3-prong $\tau$ decays with some pattern recognition
failure  or detector  inefficiency.   These were  classified into  two
categories,  $1^{\prime}$ and  $3^{\prime}$, corresponding  to 1-prong
and  3-prong  $\tau$  decays  respectively.   Candidate  $\tau$  decay
hemispheres of  type $1^{\prime}$ had to  satisfy at least  one of the
following criteria:
\begin{itemize}
\item 
one good track, with associated VD  hits, plus two
VD-ID tracks not pointing within $3^\circ$ in azimuth of a TPC sector
boundary;
\item 
one good track, plus a conversion candidate 
selected with the electron identification algorithm. 
\end{itemize}
The $3^{\prime}$ category contained all other $\tau$ decay hemispheres
failing to  pass any of  the previous classification  requirements. In
particular it  contained those  events with either  four or  five good
tracks of  which either four or  more had associated VD  hits, but not
satisfying the 5-prong class requirements.  This criteria had a higher
efficiency for 5-prong decays than  for 3-prong decays, but due to the
much  greater 3-prong  branching ratio  the $3^{\prime}$  category was
dominated  by 3-prong  decays.  Table~\ref{tab:mode_eff}  contains the
efficiencies  of  these   selection  requirements  for  the  different
exclusive  $\tau$  decay  modes  and the  inclusive  single-hemisphere
topological  selections,   as  obtained  from   simulation  and  after
corrections for observed discrepancies  between data and simulation in
the rate and reconstruction efficiency of material reinteractions.
\begin{table}[t]
\begin{center}
  \vspace{-10pt}
\begin{tabular}{l|r@{$\pm$}l|r@{$\pm$}l|r@{$\pm$}l|r@{$\pm$}l|r@{$\pm$}l|r@{$\pm$}l}
\hline
 true $\tau$ 
       & \multicolumn{2}{c}{$\tt$}      
       & \multicolumn{10}{|c}{Topology Classification}   \\ 
\cline{4-13}
 decay mode
       &\multicolumn{2}{c}{selection}
       &\multicolumn{2}{|c}{1}  
       &\multicolumn{2}{|c}{3} 
       &\multicolumn{2}{|c}{5}
       &\multicolumn{2}{|c}{$1^{\prime}$}
       &\multicolumn{2}{|c}{$3^{\prime}$}    \\
\hline                          
$\EL$               & 50.60& 0.07 & 99.95& 0.00 &  0.00& 0.00 &  0.00& 0.00 &  0.02& 0.00 &  0.02& 0.00 \\
$\MU$               & 53.31& 0.07 & 99.96& 0.00 &  0.00& 0.00 &  0.00& 0.00 &  0.01& 0.00 &  0.03& 0.00 \\
$\pi^- \nut$        & 49.69& 0.09 & 99.88& 0.01 &  0.04& 0.01 &  0.00& 0.00 &  0.00& 0.00 &  0.08& 0.01 \\
$\pi^-\pio\nut$     & 51.77& 0.06 & 97.87& 0.03 &  0.60& 0.01 &  0.00& 0.00 &  0.84& 0.02 &  0.69& 0.01 \\
$\pi^-2\pio\nut$    & 51.07& 0.11 & 95.88& 0.06 &  1.25& 0.03 &  0.00& 0.00 &  1.42& 0.04 &  1.45& 0.04 \\
$\pi^-3\pio\nut$    & 48.89& 0.25 & 94.36& 0.16 &  1.68& 0.09 &  0.00& 0.00 &  1.83& 0.10 &  2.13& 0.10 \\
$K^- \nut $         & 49.43& 0.36 & 99.90& 0.03 &  0.02& 0.02 &  0.00& 0.00 &  0.01& 0.01 &  0.06& 0.03 \\
$K^-\pio\nut$       & 51.40& 0.47 & 97.66& 0.20 &  0.85& 0.12 &  0.00& 0.00 &  0.80& 0.12 &  0.69& 0.11 \\
$K^-2\pio\nut$      & 50.42& 1.12 & 94.65& 0.71 &  2.28& 0.47 &  0.00& 0.00 &  1.39& 0.37 &  1.68& 0.40 \\
$\pi^-\kol\nut$     & 53.10& 0.48 & 99.79& 0.06 &  0.07& 0.03 &  0.00& 0.00 &  0.00& 0.00 &  0.14& 0.05 \\
$\pi^-\kol\pio\nut$ & 51.85& 0.73 & 97.32& 0.33 &  0.78& 0.18 &  0.00& 0.00 &  0.78& 0.18 &  1.11& 0.21 \\
$K^-\kol\nut$       & 54.60& 0.87 & 99.78& 0.11 &  0.11& 0.08 &  0.00& 0.00 &  0.00& 0.00 &  0.11& 0.08 \\
$K^-\kol\pio\nut$   & 52.66& 1.24 & 96.71& 0.61 &  0.94& 0.33 &  0.00& 0.00 &  0.94& 0.33 &  1.41& 0.40 \\
$\pi^-\kol\ko\nut$  & 52.82& 1.04 & 95.12& 0.62 &  3.72& 0.54 &  0.00& 0.00 &  0.08& 0.08 &  1.07& 0.30 \\
$\pi^-\kos\nut$     & 52.17& 0.48 & 94.48& 0.30 &  4.30& 0.27 &  0.00& 0.00 &  0.16& 0.05 &  1.06& 0.14 \\
$\pi^-\kos\pio\nut$ & 50.78& 0.73 & 92.64& 0.54 &  4.65& 0.43 &  0.00& 0.00 &  0.55& 0.15 &  2.16& 0.30 \\
$K^-\kos\nut$       & 52.38& 0.86 & 94.50& 0.54 &  4.42& 0.49 &  0.00& 0.00 &  0.06& 0.06 &  1.02& 0.24 \\
$K^-\kos\pio\nut$   & 51.32& 1.32 & 92.56& 0.97 &  5.01& 0.80 &  0.00& 0.00 &  0.41& 0.23 &  2.03& 0.52 \\
$\pi^-2\kos\nut$    & 46.34& 1.80 & 86.72& 1.80 & 10.45& 1.63 &  0.00& 0.00 &  0.00& 0.00 &  2.82& 0.88 \\
\hline
1-prong             & 51.42& 0.03 & 98.71& 0.01 &  0.42& 0.01 &  0.00& 0.00 &  0.44& 0.01 &  0.43& 0.01 \\
\hline
$\PPP$              & 54.71& 0.11 &  0.90& 0.03 & 90.26& 0.09 &  0.01& 0.00 &  2.10& 0.04 &  6.74& 0.07 \\
$\PPPZ$             & 53.88& 0.13 &  1.26& 0.04 & 86.39& 0.12 &  0.10& 0.01 &  3.15& 0.06 &  9.10& 0.10 \\
$\PPPZZ$            & 53.14& 0.46 &  1.37& 0.15 & 83.64& 0.46 &  0.22& 0.06 &  3.68& 0.24 & 11.09& 0.39 \\
$\PPPZZZ$           & 52.13& 1.06 &  1.46& 0.35 & 78.73& 1.20 &  0.17& 0.12 &  4.74& 0.62 & 14.90& 1.05 \\
$K^-\pi^-\pi^+\nut$ & 54.64& 0.56 &  1.03& 0.15 & 90.35& 0.45 &  0.00& 0.00 &  1.70& 0.20 &  6.92& 0.38 \\
$K^-K^+\pi^+\nut$   & 53.87& 0.90 &  2.08& 0.35 & 87.23& 0.82 &  0.00& 0.00 &  1.10& 0.26 &  9.59& 0.73 \\
\hline
3-prong             & 54.32& 0.08 &  1.06& 0.02 & 88.51& 0.07 &  0.05& 0.00 &  2.54& 0.03 &  7.84& 0.06 \\
\hline
$\PPPPP$            & 49.63& 1.19 &  0.11& 0.11 & 12.63& 1.13 & 57.52& 1.67 &  0.23& 0.16 & 29.51& 1.55 \\
$\PPPPPZ$           & 48.91& 2.23 &  0.00& 0.00 & 15.04& 2.28 & 52.85& 3.18 &  0.81& 0.57 & 31.30& 2.96 \\
\hline
5-prong             & 49.47& 1.05 &  0.09& 0.09 & 13.16& 1.01 & 56.49& 1.48 &  0.36& 0.18 & 29.90& 1.37 \\
\hline
\end{tabular}
\end{center}
\caption{\it Estimates of the selection and topology 
classification  efficiencies, in percent,
for different exclusive decay modes, as obtained from 
simulation. The efficiencies are corrected for observed 
discrepancies between data and simulation in the rate 
and reconstruction efficiency of material reinteractions. 
The quoted uncertainties are from the simulation statistics only.}
\label{tab:mode_eff}
\end{table}

Events   were   classified    according   to   both   the   hemisphere
classifications into  14 event topology  classes: 1-1, 1-$1^{\prime}$,
1-3,  1-$3^{\prime}$, 1-5,  $1^{\prime}$-$1^{\prime}$, $1^{\prime}$-3,
$1^{\prime}$-$3^{\prime}$,  $1^{\prime}$-5, 3-3,  3-$3^{\prime}$, 3-5,
$3^{\prime}$-$3^{\prime}$ and  $3^{\prime}$-5.  The class  5-5 was not
included as these  events were removed by the  multiplicity cut in the
$\tt$  preselection and  in any  case would  give a  negligible signal
contribution.  The selection of  event classes up to multiplicity five
assumes  that the  inclusive branching  ratio of  the $\tau$  to seven
charged particles is negligible.
While  inclusion  of  higher multiplicity
classes is possible,  the DELPHI sample size is  insufficient to reach
the level of precision obtained by CLEO~\cite{cleosevenlimit}.

The  various sources  of  non-$\tt$ background   are  detailed  in
Table~\ref{tab:backgrounds} for each of the event topology classes.
\begin{table}[t]
\renewcommand{\tabcolsep}{0.15pc} 
\renewcommand{\arraystretch}{1.0} 
  \vspace{-10pt}
\begin{center}
\begin{tabular}{l|c@{$\pm$}c|r|r|r|r|r|r|r|r|r|r|r|r|r|r}
\hline
{\small Source of} 
         & \multicolumn{2}{|c|}{\small All} 
         & \multicolumn{14}{c}{Event Topology}   \\ 
\cline{4-17}                        
{\small Background} &  \multicolumn{2}{c|}{\small Topologies}  
  & 1-1 & 1-$1^{\prime}$ & 1-3 & 1-$3^{\prime}$ & 1-5
  & $1^{\prime}$-$1^{\prime}$ & $1^{\prime}$-3
  & $1^{\prime}$-$3^{\prime}$ & $1^{\prime}$-5
  & 3-3 & 3-$3^{\prime}$ & 3-5 & $3^{\prime}$-$3^{\prime}$ 
  & $3^{\prime}$-5 \\
\hline                                                                                                                    
$\mm$      & 0.11 & 0.01 & 0.19 & 0.06 & 0.00 & 0.07 & 0.00 & 0.00 & 0.00 & 0.00 & 0.00 & 0.00 & 0.00 & 0.00 & 0.00 & 0.00\\
$\ee$      & 0.40 & 0.07 & 0.65 & 0.00 & 0.01 & 0.00 & 0.00 & 0.00 & 0.00 & 0.00 & 0.00 & 0.00 & 0.00 & 0.00 & 0.00 & 0.00\\
$\qq$      & 0.29 & 0.01 & 0.03 & 0.00 & 0.32 & 3.39 & 1.01 & 0.00 & 1.16 & 0.00 & 0.00 & 1.61 & 14.8 & 21.2 & 77.0 & 0.00\\
$\ee\ee$   & 0.27 & 0.03 & 0.49 & 0.00 & 0.00 & 0.00 & 0.00 & 0.00 & 0.00 & 0.00 & 0.00 & 0.00 & 0.00 & 0.00 & 0.00 & 0.00\\
$\ee\mm$   & 0.10 & 0.01 & 0.16 & 0.00 & 0.01 & 0.00 & 0.00 & 0.00 & 0.00 & 0.00 & 0.00 & 0.00 & 0.33 & 0.00 & 0.00 & 0.00\\
$\ee\tt$   & 0.27 & 0.03 & 0.35 & 3.91 & 0.23 & 0.76 & 0.00 & 21.1 & 0.97 & 2.88 & 0.00 & 0.20 & 0.60 & 0.00 & 0.45 & 0.00\\
$\ee\qq$   & 0.02 & 0.01 & 0.03 & 0.00 & 0.03 & 0.44 & 0.00 & 0.00 & 0.00 & 0.00 & 0.00 & 0.00 & 0.00 & 0.00 & 0.00 & 0.00\\
cosmic rays& 0.05 & 0.01 & 0.07 & 0.01 & 0.00 & 0.00 & 0.00 & 0.00 & 0.00 & 0.00 & 0.00 & 0.00 & 0.00 & 0.00 & 0.00 & 0.00\\
\hline                                                                                                                    
Total      & 1.51 & 0.10 & 1.97 & 3.98 & 0.60 & 4.66 & 1.01 & 21.1 & 2.13 & 2.88 & 0.00 & 1.81 & 15.7 & 21.2 & 77.4 & 0.00\\
\hline                                    
\end{tabular}                             
\end{center}                              
\caption{\it  Fractional non-$\tt$ backgrounds (in \%) in the $\tt$ 
sample and the different event topology classes.
The classes 1$^{~\!\!\prime}$-5 and 3$^{~\!\!\prime}$-5 have 
very small populations and the estimated 
background is zero due to statistical fluctuations in the simulation.}
\label{tab:backgrounds}                   
\end{table}                               

\section{The fit and systematics}
\label{sec:fit}
\subsection{Fitting procedure}
\label{sec:fitmethod}
A reweighting  technique was  used to take  into account  the observed
data/simulation   discrepancies  in   the  rates   and  reconstruction
efficiencies  of   nuclear  reinteractions,  photon   conversions  and
$\delta$-rays.   Each   secondary  or  tertiary  particle   $i$  in  a
simulated~$\tt$  event~$k$  was   given  a  weight~$W^{part}_i$.   For
particles  interacting  with the  detector  material  this weight  was
obtained        from       the       studies        described       in
Section~\ref{sec:reconstruction}        and        summarised       in
Tables~\ref{tab:conversion_corrections}
and~\ref{tab:reinteraction_corrections}.    These   weights   were   a
function of the radius and type of the reinteraction, as follows:
\begin{itemize}
\item
reconstructed converted photons: 
$W^{part}_i = A^{^{conv}}_{_{rec}}({\mathrm r})$; 
\item
unreconstructed converted photons: 
$W^{part}_i = A^{^{conv}}_{_{unrec}}({\mathrm r})$; 
\item
reconstructed inelastic nuclear reinteractions: 
$W^{part}_i = A^{^{nucl}}_{_{rec}}({\mathrm r})$;
\item
unreconstructed inelastic nuclear reinteractions: 
$W^{part}_i = A^{^{nucl}}_{_{unrec}}({\mathrm r})$; 
\item
$\delta$-rays:
$W^{part}_i = X^{^{EM}}_{d}/X^{^{EM}}_{s}({\mathrm r})$;
\item
bremsstrahlung emitting electrons:
$W^{part}_i = X^{^{EM}}_{d}/X^{^{EM}}_{s}({\mathrm r})$;
\item
elastic nuclear reinteractions:
$W^{part}_i = X^{^{nucl}}_{d}/X^{^{nucl}}_{s}({\mathrm r})$.
\end{itemize}
It was also  necessary to reweight particles which  did not reinteract
so as to maintain the  internal normalisation of the simulation sample
with  respect to  parameters such  as angular  distributions, momentum
distributions, or  the $\tau$ exclusive branching  ratios.  Failure to
achieve this  could lead to biases  within the simulation  sample as a
function  of  photon  multiplicities, charged  hadron  multiplicities,
angular regions with different amounts  of material, or could bias the
preselection  of the  $\tt$ sample  with respect  to certain  types of
decay     mode.     Non-interacting     photons    were     given    a
weight~$W^{part}_i$~given~by
\begin{equation}
W^{part}_i = 
  \frac{N^{^\gamma}-\!\!\!\!\displaystyle\sum_{j \in \mathrm{int. photons}} 
  \!\!\!\!\!\!\!\! W^{part}_j}{N^{^\gamma}-N^\gamma_{int}},
\end{equation}
where $N^\gamma$  is the  total number of  photons produced  by $\tau$
decays or  by interactions with  material of $\tau$ decay  products in
the simulation sample. $N^\gamma_{int}$  is the number of photons with
a material reinteraction. The sum  is over all photons with a material
reinteraction. To maintain the  angular and momentum distributions the
above weight calculation was  performed separately for individual bins
in  a three-dimensional  space of  the photon  polar  angle, azimuthal
angle  and  momentum.   A  similar  procedure  was  used  to  reweight
non-interacting hadrons, particles not emitting a $\delta$-ray (a very
small correction) and electrons without any bremsstrahlung emission.

With every  secondary and tertiary  particle $i$ in a  simulated event
$k$ given a weight $W^{part}_i$, the weight $W^{event}_k$ of the event
was given by the product of all these weights:
\begin{equation}
W^{event}_k = \prod_{i \in k} W^{part}_i.
\end{equation}

The numbers of selected events  in each event topology class are shown
in Table~\ref{tab:class_numbers}.   A maximum likelihood  fit assuming
Poissonian probabilities  was performed  to the reweighted  numbers of
events estimated using Eqn.~\ref{eqn:classpred2}.  The number of $\tt$
events ($N_{\tau\tau}$ in Eqn.~\ref{eqn:classpred2}) and the branching
ratios $B_1$ and $B_5$ were allowed to vary in the fit while $B_3$ was
constrained by the relation $B_1+B_3+B_5=1$.  The output of the fit is
also shown in Table~\ref{tab:class_numbers}.   The results of this fit
were:  
$\Bone   =(85.316\pm0.093)\%$;  
$\Bthree  =(14.569\pm0.093)\%$;
$\Bfive =(0.115\pm0.013)\%$, 
where only the statistical error from the
fit is quoted.  An estimate of  the consistency of the fit was made by
calculating a  $\chi^2$. This took  into account only  the statistical
uncertainties. It gave a $\chi^2/$n.d.f. of $16.4/11$, indicating good
consistency. The contribution to the $\chi^2$ from each class is shown
in Table~\ref{tab:class_numbers}.
\begin{table}[t]
\begin{center}
\renewcommand{\tabcolsep}{0.2pc} 
  \vspace{-10pt}
\begin{tabular}{@{~}l@{$-$}l@{~}|r|r|@{~~}r@{~~~}}
\hline
 \multicolumn{2}{c|}{Class}    & Observed & Fit Output &$\chi^2$ \\
\hline 
 1  &  1                       &    56219 &   56149.0  &  0.1  \\
 1  &  $1^{\prime}$            &      858 &     871.0  &  0.2  \\
 1  &  3                       &    18681 &   18813.5  &  0.9  \\
 1  &  $3^{\prime}$            &     2350 &    2331.5  &  0.1  \\
 1  &  5                       &       94 &      95.6  &  0.0  \\
 $1^{\prime}$ &  $1^{\prime}$  &        4 &       3.9  &  0.0  \\
 $1^{\prime}$ &  3             &      131 &     134.8  &  0.1  \\
 $1^{\prime}$ &  $3^{\prime}$  &       16 &      14.7  &  0.1  \\
 $1^{\prime}$ &  5             &        0 &       1.2  &  1.2  \\
 3  &  3                       &     1481 &    1451.5  &  0.6  \\
 3  &  $3^{\prime}$            &      409 &     357.4  &  7.4  \\
 3  &  5                       &       17 &      13.8  &  0.7  \\
 $3^{\prime}$ &  $3^{\prime}$  &       76 &      97.7  &  4.8  \\
 $3^{\prime}$ &  5             &        1 &       1.5  &  0.2  \\
\hline                         
\multicolumn{2}{c|}{All}       &    80337 &   80337.1  & 16.4  \\
\hline 
\end{tabular}
\end{center}
\caption{\it The second and third columns contain the number of 
observed events in each class and number predicted from the
maximum likelihood fit of the $\tau$ topological branching ratios.
The fourth column contains the $\chi^2$ contribution for each class
to the overall $\chi^2$.}
\label{tab:class_numbers}
\end{table}

\subsection{Systematics}
\label{sec:systematics}
In general, the systematic  uncertainties on the topological branching
ratios due to any  particular effect were estimated simultaneously for
$B_1$, $B_3$ and $B_5$ by  repeating the analysis, including the $\tt$
preselection, after modifying the relevant variable in the simulation.
This  accounted  for  correlations  in  the  systematic  uncertainties
between   the   different  branching   ratios   and  correlations   in
efficiencies and  backgrounds between the different  event classes and
the $\tt$ preselection.

\subsubsection{Preselection}
\label{sec:systematics_preselection}
The main  systematic effects  of the $\tt$  selection criteria  on the
result can arise through the mis-calibration of the quantities used in
the selection. These quantities can be classified into cuts related to
energy  or momentum  measurements,  such as  $E_{rad}$, $p_{rad}$  and
$E_{vis}$,  or  those  cuts  related  to  multiplicity,  such  as  the
isolation angle, which can be incorrectly estimated if extra secondary
particles are produced in  reinteractions with detector material. This
second  effect  is  taken  account   of  by  the  systematics  in  the
reweighting    of   the    secondary   interactions,    discussed   in
Section~\ref{sec:systematics_reinteractions}.   A  separate systematic
uncertainty  is included  for the  energy and  momentum  scales below.
Other  effects such  as  tracking efficiency,  trigger efficiency  and
uncertainties  on the $\tau$  exclusive branching  ratios can  have an
effect  on  the  $\tt$  preselection  efficiency.   These  sources  of
systematic  uncertainty  are  discussed  below,  and  in  general  the
systematic uncertainties on the topological branching ratios take into
account effects in the preselection.

Any  remnant discrepancies  due  to $\tt$  preselection criteria  were
checked by studying the agreement  between data and simulation for the
distributions  of  the  $\tt$  selection  variable  for  each  of  the
different  event  topology  classes.   With  the  corrections  to  the
background levels  discussed in Sections~\ref{sec:backgrounds}
and~\ref{sec:reconstruction}, good agreement  was found, in
particular in the regions of the cuts.

The uncertainties  due to the backgrounds from  non-$\tt$ sources were
estimated   by  varying   the  background   normalisations   by  their
uncertainties         obtained         as         described         in
Section~\ref{sec:tautauselection},      and     are      listed     in
Table~\ref{tab:uncertainties}.  The distribution of the neural network
variable used for $\qq$ rejection (see Fig.~\ref{fig:nnqq}) displays a
15\%  excess  in the  data  compared  with  simulation in  the  region
[0.05;0.8]  of  the  output  neuron  distribution.   This  region  was
dominated by  $\tt$ events and  was studied to  see if there  were any
discrepancies in the input variables, and by checking the multiplicity
distributions  of  the  events.  No  discrepancies  were  found.   The
systematic  uncertainty  was estimated  by  rescaling  the numbers  of
rejected signal events in all classes by~15\%.

\subsubsection{Tracking}
\label{sec:systematics_tracking}
Within  the  angular  acceptance  of  this analysis,  there  are  four
tracking detectors which can contribute to the track reconstruction of
all charged particles:  the VD, ID, TPC and  OD. The reconstruction of
``good''  tracks  is strongly  dependent  on  the  TPC with  its  full
three-dimensional readout.  The redundancy in the track reconstruction
obtained  by the  inclusion of  $\tau$ decay  hemispheres with  only a
VD-ID track in the $\tt$  sample and 1-prong subsample reduces greatly
the sensitivity  of the measurement  to inefficiencies within  a given
subdetector, and allows direct cross-checks to be made.

In simulation,  for non-interacting particles, the  efficiency for the
TPC to reconstruct an isolated charged particle which passes through a
sensitive gas volume far from  dead regions is 99.95\%.  However, even
with the  redundancy of the different  tracking subdetector components
in DELPHI,  to measure this in  data directly is  difficult because of
material reinteractions between different subdetectors which can cause
the   particle  to  be   lost  before   entering  the   TPC  sensitive
volume. These effects tend to reduce the measured efficiency.  For the
data, the efficiency of the  TPC in its sensitive regions
was estimated using the redundancy of
the tracking system to be $(99.35\pm0.05)\%$.  The identical procedure
was   applied   to  the   simulation,   yielding   an  efficiency   of
$(99.31\pm0.01)\%$, in excellent agreement with the data, and implying
that  the  modelling of  the  TPC  efficiency  in the  simulation  was
accurate. The true inefficiency in  simulation is a factor of 12 lower
than the  inefficiency estimated by  this method.  The  uncertainty on
the  TPC reconstruction  efficiency in  the TPC  sensitive  region was
taken  conservatively to  be $0.05\%$  to account  for  any systematic
effects in its estimation.

The TPC efficiency  for isolated tracks in the  sector boundary region
was studied using inclusive  low multiplicity events but excluding the
$\tt$ events.   This sample consists of radiative  dilepton events and
the high energy  part of the two-photon leptonic  event spectrum.  The
number of reconstructed tracks in the TPC boundary region was compared
to that expected  by normalisation of the sensitive  region of the TPC
sectors, in bins of momentum.   The loss of tracks was compatible with
zero below about  5~GeV/$c$, rising to 3.5\% at  45~GeV/$c$.  Data and
simulation agree well in  their behaviour.  The inefficiency estimated
from data  for the $\tt$ momentum  distribution was $(2.73\pm0.04)\%$,
compatible  with  the  result  in  simulated $\tt$  events.   The  TPC
efficiency was varied  by throwing away tracks containing  TPC hits in
simulation and repeating the analysis, including the $\tt$ preselection
stage.  The error  was  scaled by  a  factor two  to  account for  any
systematic effects in the procedure.

The attachment  of VD hits  to a track  is the main criterion  used in
this analysis  to determine the  multiplicity of a $\tau$  decay.  The
association efficiency and mis-association  probability of a VD hit to
be    attached    to   a    charged    particle    track   has    been
studied~\cite{btagging}  on large  samples of  $\eeha$ events,  and on
isolated  topologies  such  as  $\mm$  final  states.   These  studies
indicate that  the efficiency in  simulation and data agree  to within
$\pm2\%$. The  associated systematic uncertainties  were obtained from
the  shifts in  the results  observed  when randomly  removing 2\%  of
associated  VD hits  and repeating  the analysis,  including  the $\tt$
preselection stage.

Within the $\tau$  sample itself the rate of  association of different
subdetectors to a reconstructed charged particle track was studied for
different $\tau$  decay topologies.  In candidate  1-prong decays, the
hit association probabilities  for the VD, ID and  OD were studied for
tracks  which  had  an  associated  TPC  track  segment.
The probabilities  obtained  in data  and
simulation    were    compared    and   the    relative    differences
calculated. These differences were 0.0\% for the VD, +2.3\% for the ID
and $-$1.2\%  for the  OD. The equivalent  numbers for  3-prong decays
were  $-$0.7\%,  +2.5\% and  +0.6\%  respectively.  
The results for  the ID
were unchanged if in addition the  VD was required to be associated to
a track, and  vice-versa.  The small fraction of  tracks without a TPC
track  segment showed  a level  of  agreement for  the proportions  of
tracks with  different subdetectors which  was better than 3\%  in all
cases.  

In simulation, the preselection efficiency was found to be independent
of the combination of the subdetectors  attached to a track at a level
below~0.1\%.  Given  the   observed  discrepancies  between  data  and
simulation,  any effect  on the  preselection efficiency  due  to this
source was~${\cal{O}}(10^{-5})$.   The uncertainties arising  from the
probability of  including an ID  hit on a  track were estimated  to be
$2.2\times10^{-5}$   for   both   $B_1$   and  $B_3$   with   a   full
anticorrelation  and  has been  included  in  the tracking  systematic
uncertainty.  The analogous uncertainty for the OD was negligible.

It was  observed that  the level of  Bhabha background  was correlated
with  the  existence  of an  OD  hit  on  a  track.  This was  due  to
bremsstrahlung, and resolution effects  in the $p_{rad}$ variable. The
level  of  agreement observed  between  the  data  and the  simulation
implied that the Bhabha  background was consistent with the estimation
made in Section~\ref{sec:backgrounds}.

\begin{figure}[t]
  \begin{center}
  \vspace{-25pt}
\epsfig{figure=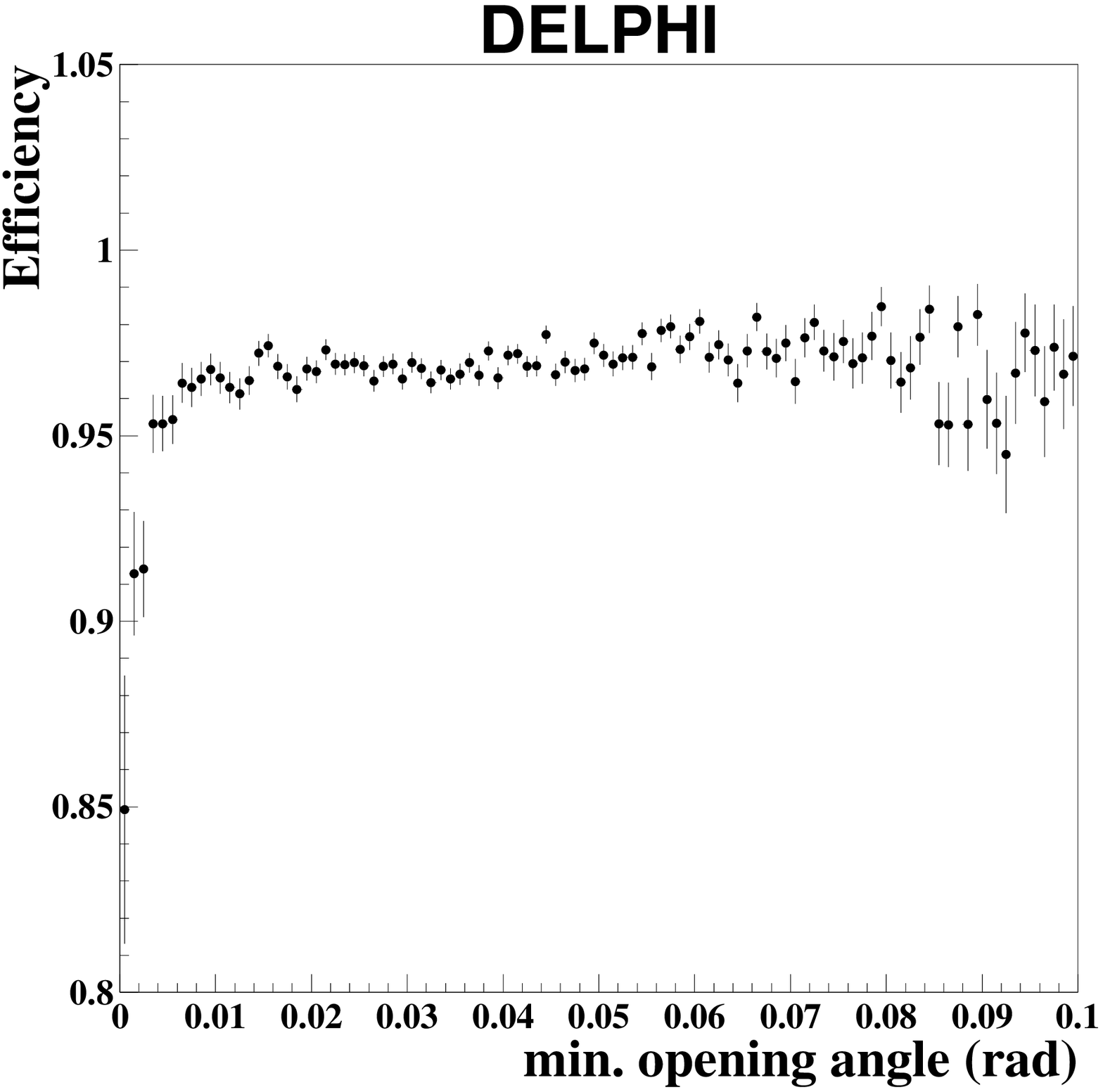,width=0.49\linewidth,height=8cm}
\put(-50,170){\bf\huge a}    
\hfill
\epsfig{figure=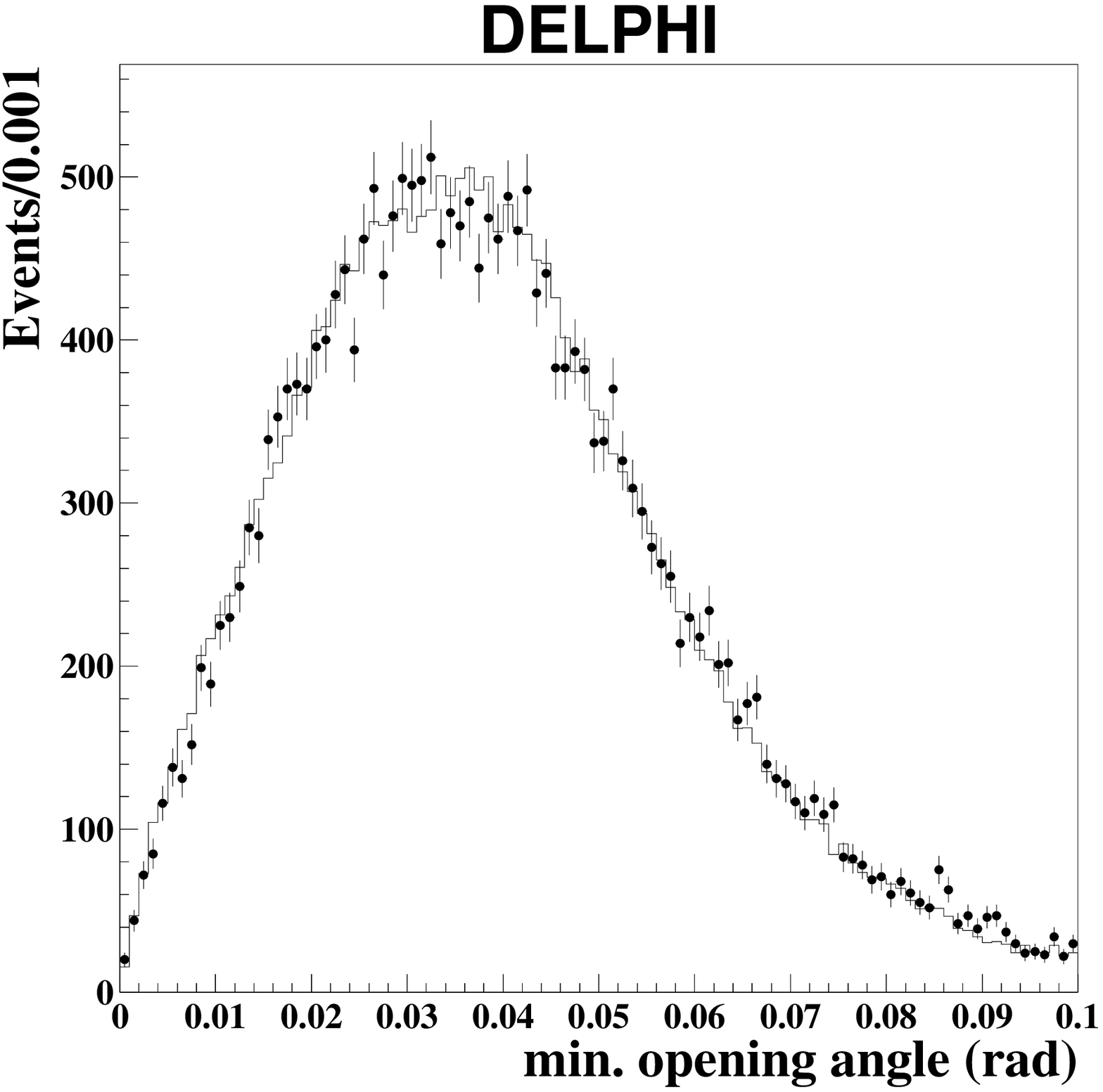,width=0.49\linewidth,height=8cm} 
\put(-50,170){\bf\huge b}    
    \caption{\it a) The selection efficiency in simulation for candidate
             3-prong decays as a function
             of the minimum angle in three dimensions
             between any two charged particles.
             b)~The~distribution of the
             minimum angle in three dimensions
             between any two charged particles
             in candidate 3-prong $\tau$ decays. 
             Dots are data,  line is simulation.} 
    \label{fig:opening_angle_3}
  \end{center}
\end{figure}
\begin{figure}[t]
  \begin{center}
  \vspace{-25pt}
    \epsfig{figure=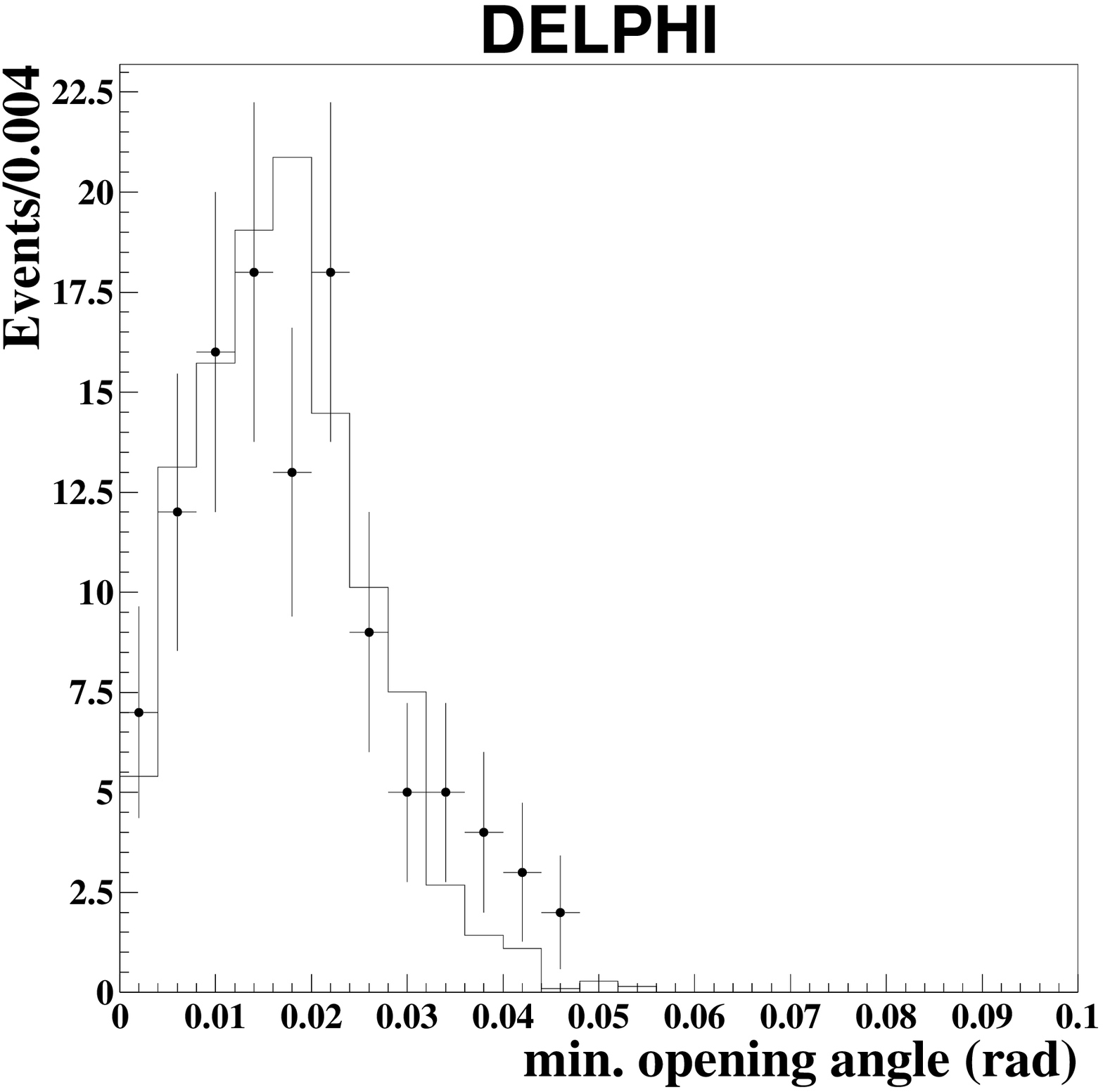,width=0.6\linewidth,height=8cm} 
    \caption{\it The minimum angle in three dimensions
             between any two charged particles
             in candidate 5-prong $\tau$ decays.  
             Dots are data,  line is simulation.
             } 
    \label{fig:opening_angle_5}
  \end{center}
\end{figure}
The   two-track  resolution  has   been  studied   by  data-simulation
comparison of  the minimum opening  angle in three  dimensions between
tracks  in  $\tau$  decay   hemispheres  with  more  than  one  track.
Fig.~\ref{fig:opening_angle_3}a shows the  efficiency in simulation to
reconstruct  a 3-prong  $\tau$  decay  as a  function  of the  minimum
opening angle.  It  is flat within 1\% except for  a fall-off of about
5\%  in reconstruction efficiency  for a  minimum opening  angle below
3~mrad.   As is  visible in  the distribution  of the  minimum opening
angle    in    candidate    3-prong    $\tau$    decays    shown    in
Fig.~\ref{fig:opening_angle_3}b,  less  than  1\%  of  3-prong  $\tau$
decays  lie in  this region  and  data and  simulation are  compatible
within  the  statistical   precision.  A  systematic  uncertainty  was
attributed to  the two-track resolution  by varying the  efficiency in
the region with minimum opening  angle below 3~mrad by the uncertainty
allowed from  the statistical uncertainty  of the data.  There  was no
observable  fall-off  in efficiency  for  low  values  of the  minimum
differences  between the  azimuthal or  polar  angles of  tracks in  a
3-prong  hemisphere.   The   minimum  opening  angle  distribution  in
candidate       5-prong       $\tau$       decays       is       shown
in~Fig.~\ref{fig:opening_angle_5}.

\begin{figure}[t]
  \begin{center}
  \vspace{-25pt}
\epsfig{figure=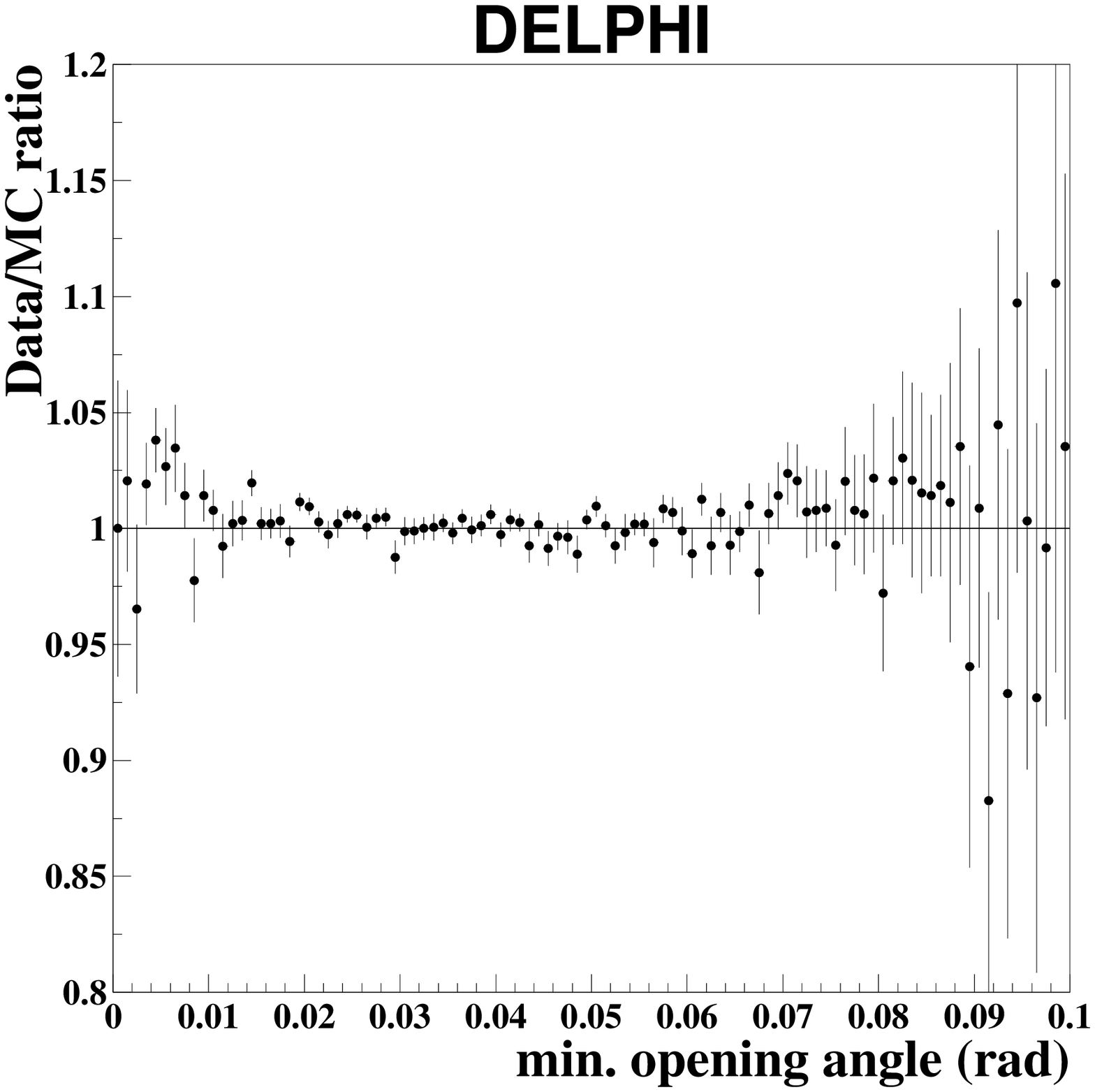,width=0.6\linewidth,height=8cm} 
    \caption{\it For candidate 3-prong $\tau$-decays,
    the ratio of data to simulation of the fraction of candidate 
    decays where three tracks had associated TPC hits, 
    as a function of the minimum opening angle.}
    \label{fig:tpc_twotrack}
  \end{center}
\end{figure}
The association efficiency of the TPC in 3-prong $\tau$ decays and its
dependence on the minimum opening  angle were studied by comparing the
rate  of   candidate  $\tau$  decays  containing   three  tracks  with
associated  TPC hits  with the  rate of  decays containing  either two
tracks with TPC hits plus one track without TPC hits or one track with
TPC hits plus two tracks without TPC hits.  
Fig.~\ref{fig:tpc_twotrack}
shows  the data  over simulation  ratio of  the fraction  of candidate
3-prong  $\tau$ decays  containing  three tracks  with associated  TPC
hits,  as  a  function  of   the  minimum  opening  angle.   It  shows
compatibility with unity in all regions except for a small discrepancy
of about +3\% for minimum opening angles in the region of 4 to 7 mrad.
The  average  of   the  ratio  for  all  minimum   opening  angles  is
$1.0016\pm0.0008$,   consistent  with   unity   within  two   standard
deviations.  
The difference from unity can be directly related to the
efficiency for a charged particle track to have associated TPC hits in
3-prong $\tau$ decays.  The uncertainties on the topological branching
ratios were estimated by  varying the probability in simulated 3-prong
$\tau$ decays  to have TPC hits  associated to a track  by $1.6 \times
10^{-3}$ and  repeating the analysis.  The same  procedure was carried
out for 5-prong decays. 

Fig.~\ref{fig:opening_angle_3}b  shows some  differences  between data
and simulation in  the region between~3 and 15~mrad,
with the simulation tending to lie
slightly above the data.  This  can be  due to  the decay
modelling,  for  which  the  attendant  systematic  uncertainties  are
discussed  in  Section~\ref{sec:systematics_br}. However 
track  reconstruction effects cannot  be excluded. While systematics
in the TPC association efficiency are accounted for above, 
the deviation from  unity of  the ratio
shown in  Fig~\ref{fig:tpc_twotrack} can also give an  estimate of the
rate  of track reconstruction  without the  TPC,  in particular of  VD-ID
tracks.  An  upper estimate of the  magnitude of this  effect
for close tracks
was derived  by integrating up to 15~mrad the deviations from  unity of  this ratio
times the number of  events in a given bin of minimum opening  angle.
This procedure
gave an  uncertainty of  $1.2 \times10^{-4}$ on  $B_3$, 
together with  uncertainties   on  $B_1$  
and  $B_5$ induced via  the correlations in the fit. This accounted for
potential differences between data and simulation in the
VD-ID track reconstruction in the small minimum opening angle
region of 3-prong $\tau$ decays.

The reconstructed charge in a  $\tau$ decay hemisphere was studied for
different  event and  decay topologies.  The  charge was  not used  to
classify $\tau$  decays or  select $\tt$ events,  so this  provided an
indirect cross-check of the track reconstruction.  Use was made of the
constraint that  the two $\tau$-decay hemispheres  arose from $\tau$'s
of opposite charge.  For events in the data belonging to the 1-1 event
class, an estimated (99.88$\pm$0.01)\% of $\tau$ decays had the charge
correctly   reconstructed.    The   rate   in   the   simulation   was
(99.91$\pm$0.01)\%.   For 3-prong  decays in  data, (98.57$\pm$0.09)\%
had the correctly signed unit charge, compared with (98.60$\pm$0.03)\%
in simulation, while (0.54$\pm$0.06)\% had  a charge of three with the
correct sign  as compared  with (0.52$\pm$0.02)\% in  simulation.  The
remainder  had  the  wrongly  signed  charge.  In  5-prong  decays  an
estimated (90.5$\pm$0.7)\% had the correctly reconstructed unit charge
in data compared with (89.7$\pm$0.3)\% in simulation.

\subsubsection{Reinteractions and {\boldmath $\kos$} reconstruction}
\label{sec:systematics_reinteractions}
Uncertainties from the photon conversion reconstruction were estimated
by  varying by  their  uncertainties the  correction  factors for  the
reconstructed    and    unreconstructed    conversions,    given    in
Table~\ref{tab:conversion_corrections}, which  were obtained from data
test  samples   as  described  in   Section~\ref{sec:conversions}.   A
contribution  for the  uncertainty in  the incident  photon  rate 
in the dileptonic test samples was
included.    The  resultant   uncertainties  are   dominated   by  the
contribution from the unreconstructed conversions.  A similar approach
was taken for the  nuclear reinteractions.  Here again the uncertainty
arising from  the unreconstructed nuclear reinteractions  was the most
significant contribution.

Both the  photon conversion and hadronic  reinteraction errors contain
contributions  to account  for the  uncertainty  on the  rate of  fake
reconstructions which cause the topology to be misidentified.

The 1-prong  $\tau$ decay class  has a number of  different subclasses
which cover the different  reconstruction possibilities for an elastic
nuclear  reinteraction  in  a   1-prong  hadronic  $\tau$  decay.   In
simulation, of  the 3.3\%  of hadrons in  $\tau$ decays  undergoing an
elastic scatter before the  TPC sensitive volume 99.4\% were correctly
attributed  to   the  1-prong  class,  leading   to  a  classification
inefficiency of $2.0\times10^{-4}$ for 1-prong hadronic $\tau$ decays.
A  cross-check of  the elastic  scatters  was performed  using a  kink
algorithm which attempted  to link up pairs of  tracks consistent with
both  tracks  having  been  produced  by  a  single  charged  particle
experiencing a large scatter  in detector material.  Comparison of the
rates in data  and simulation showed agreement within  20\%, which was
taken as  a relative uncertainty on  the effect.  The  effects of fake
kink reconstructions by the algorithm had a negligible effect.

The  uncertainty   attributed  to  the   electron  identification  had
contributions   from  the  uncertainties   on  the   efficiencies  for
conversion rejection  and for rejection of $\pi^0$  Dalitz decays, and
from the probability  of misidentifying a hadron as  an electron.  The
study  of the  electron  identification algorithm  and its  systematic
uncertainties is described in Section~\ref{sec:conversions}.

The $\delta$-ray rate is proportional to the mean atomic number
Z of the material, while the factors
which are applied correct for the
number of  radiation lengths, which is approximately proportional 
to the  mean Z$^2$.
This gives an ambiguity  in the correction for $\delta$-rays depending
on whether it is due to the  wrong mean Z, or to the wrong quantity of
material, but with the correct  mean Z.  The correction factor used in
the analysis, $X^{EM}_d/X^{EM}_s$, assumes  the latter case.  The former
case  would  imply a  factor  of  $(X^{EM}_d/X^{EM}_s)^{1/2}$.  This  was
applied and  the observed variations on the branching ratios  taken as
systematic uncertainties.

\enlargethispage*{5mm}
Only  the   fraction  ($\approx$5\%)  of   $\kos\to\pi^+\pi^-$  decays
occuring inside the VD,  and which the $\kos$ reconstruction algorithm
failed to  identify, gave fake  charged primary $\tau$  decay products
and hence  the incorrect topology  assignment.  For the  region inside
the  outer layer  of the  VD, the  rate of  successfully reconstructed
$\kos$'s  was $0.91\pm0.09$ times  the simulation  prediction, showing
good  consistency.   The  quoted  error  is  purely  statistical.   By
comparison  of  data  and   simulation  for  candidate  $\kos$  mesons
reconstructed as decaying beyond the VD, the reconstruction efficiency
and relative  rate of photon conversions misidentified  as $\kos$ were
estimated to  agree within 15\%.  The  $\kos$ reconstruction algorithm
can reconstruct a fake $\kos$  from two primary $\tau$ decay particles
thus reducing the $\tau$ decay charged multiplicity by two. Simulation
studies indicate that the scale  of this effect was $5 \times 10^{-5}$
of the 3-prong rate.  The  related systematic uncertainty was taken to
be equal to the size of the effect.

\subsubsection{Exclusive branching ratios 
               and {\boldmath $\tau$} decay modelling}
\label{sec:systematics_br}
The exclusive  branching ratios  were varied within  the uncertainties
quoted in the Particle  Data Listings~\cite{pdg2000}.  To take account
of any  hidden correlations the quoted uncertainties  have been scaled
up by  a factor of 1.5.   The largest single contribution  is from the
decay  mode $\PPPZZZ$  with a  high  $\pio$ multiplicity  and a  large
relative  uncertainty on  the  branching ratio.  The uncertainty  also
included  a  small contribution  due  to  the  decay modes  $\tau^-\to
K^-\pi^-\pi^+\pio\nut$ and  $\tau^-\to K^-K^+\pi^-\pio\nut$ which were
not included in the simulation.

The uncertainties associated with the  modelling of the 3- and 5-prong
decays  were  estimated by  correcting  the  efficiencies taking  into
account  differences   between  data  and   simulated  invariant  mass
distributions. In addition, the hadronic structure of the $3\pi$ final
state was  varied between  the default TAUOLA~\cite{tauola}  model and
that obtained in the DELPHI analysis of the $3\pi$ structure in $\tau$
decays~\cite{delphithreepi}.    For  the   $3\pi\pio$   structure  the
parameterisation of  Model~1 of~\cite{cleostruct}  was used and,  as a
cross-check,     the     parameterisation     of    $3\pi\pio$     used
in~\cite{delphithreepi} was used to  reweight the distributions of the
minimum opening angle.

\subsubsection{Trigger, energy scale, {\boldmath $\tau$} 
               polarisation and simulation statistics}
\label{sec:systematics_trigger}
The trigger  efficiency for $\tt$ final  states was $(99.98\pm0.01)\%$
for events within the  polar angle acceptance.  Studies indicated that
the inefficiency  was due to  events where both $\tau$'s  decayed via
the   $\tau\to\mu\nu\nu$  mode~\cite{delphileptonicbr99}.    
It  was   assumed  that   the  full
inefficiency of $(2\pm1)\times 10^{-4}$ was contained in the 1-1 event
class and the  efficiency of this class and  of the $\tt$ preselection
were modified accordingly.   The associated systematic uncertainty was
obtained by varying the inefficiency by its error.

The energy and  momentum scales and resolution can  affect many of the
quantities  used  in  the   analysis,  such  as  the  $\tt$  selection
variables,  invariant  masses  or   the  thrust  (used  in  the  $\qq$
rejection).   The momentum  scale was  varied by  0.2\%,  the electron
energy  by 0.5\%  and  the neutral  electromagnetic  energy by  0.2\%.
These  variations were  obtained in  a  study carried  out for  the
$\tau$~polarisation    measurement~\cite{delphitaupolarisation}.  The
associated systematic 
uncertainty on the topological branching ratio results
was small.

The  1-prong selection  efficiency  has a  slight  sensitivity to  the
average   $\tau$  polarisation  because   of  acceptance   effects  in
$\tau\to\pi\nut,K\nut$  decays due to  the $p_{rad}$  cut used  in the
$\tt$ selection.   The analysis used  the result and  uncertainty from
the           DELPHI            analysis           on           $\tau$
polarisation~\cite{delphitaupolarisation}.

The systematic  uncertainty due  to the limited  simulation statistics
was also included.

\subsubsection{Other cross-checks and summary}
\label{sec:systematics_other}
A  cross-check of  the fitting  procedure was  performed by  using the
simulation itself  as input to  the fit; the  results of the  fit were
compared with  the input branching ratios and  agreement was observed.
This was repeated with different input branching ratios, again showing
good  agreement.   The  reweighting  procedure  was,  where  possible,
cross-checked by  direct calculation of the expected  variation in the
measured  branching  ratios when  applying  the  weights  for a  given
effect.

In  the fit  the  main contributions  to  the $\chi^2$  come from  the
3-$3^{\prime}$  and $3^{\prime}$-$3^{\prime}$  classes.   Both contain
significant  background from $\ee\to\qq$  events.  Removing  these two
classes from the  fit had an almost negligible  effect on the results,
within the bounds of  expected statistical fluctuations. No systematic
uncertainty was ascribed to this effect.

Other  cross-checks were  performed, including  fitting  the branching
ratios  hemisphere by  hemisphere rather  than event  by  event, using
Eqn.~\ref{eqn:classpred}. For  this fit it was  practical to subdivide
the sample  even further into  hemisphere subclasses dependent  on the
number of VD-ID-TPC(-OD) tracks,  VD-ID tracks, number of conversions,
nuclear reinteractions, etc.  The  branching ratios obtained with this
approach  were  in excellent  agreement  with  those  obtained in  the
event-by-event fit.

The  fits  were  performed  for  each year's  data  separately.   Good
agreement was found for the results in the different years.

The   different    systematic   uncertainties   are    summarised   in
Table~\ref{tab:uncertainties}. The systematic uncertainties are signed
so as to give the correlation between the different branching ratios.
\begin{table}[t]
  \begin{center}
  \vspace{-10pt}
    \begin{tabular}[center]{l|r|r|r}
\hline
Source of systematic     & 1-prong  & 3-prong  & 5-prong  \\
\hline
Dilepton background      &  110     & $-$109   &    $-$1  \\
Cosmic ray background    &    5     &   $-$5   &    $<$1  \\
Four-fermion background  &   42     &  $-$41   &    $-$1  \\
$Z\to\qq$ background     &   25     &  $-$24   &    $-$1  \\
Neural Network $\qq$ rejection
                         &   50     &  $-$48   &    $-$5  \\
Tracking                 &  157     & $-$152   &   $-$16  \\
VD efficiency            &   55     &  $-$60   &      +6  \\
Conversions              &  126     & $-$121   &    $-$8  \\
Inelastic Nucl. reinteractions
                         &   90     &  $-$80   &   $-$10  \\
Elastic Nucl. reinteractions
                         &   24     &  $-$24   &    $-$2  \\
Electron identification  &  104     &  $-$97   &    $-$7  \\
$\delta$-ray weights     &    8     &   $-$8   &    $<$1  \\
$\kos$ reconstruction    &    5     &   $-$5   &    $<$1  \\
Exclusive BRs            &  228     & $-$204   &   $-$44  \\
3-prong decay modelling  &  116     & $-$121   &     +10  \\
Trigger                  &   15     &  $-$15   &    $<$1  \\
E and p scales           &   19     &  $-$20   &      +1  \\
$\tau$ polarisation      &   18     &  $-$19   &      +1  \\
Simulation statistics    &  310     & $-$310   &     +31  \\
\hline                                             
Total systematic         &  492     &    477   &      59  \\  
Statistical              &  929     &    929   &     126  \\  
\hline
    \end{tabular}
    \caption{\it Contributions in units of ${\mathit{10^{-6}}}$ 
to the systematic 
uncertainties on $B_1$, $B_3$ and~$B_5$. The uncertainties are signed
to show the correlation between the different branching ratios. The
1-prong errors are always assumed positive.}
    \label{tab:uncertainties}
  \end{center}
\end{table}

\section{Conclusions}
\label{sec:conclusions}
\noindent
The measurements  made of the 1, 3  and 5-prong topological
branching ratios were
\[\begin{array}{ll}
\Bone      & = (85.316\pm0.093_{stat}\pm0.049_{sys})\%,\\
\Bthree    & = (14.569\pm0.093_{stat}\pm0.048_{sys})\%,\\
\Bfive     & = (0.115\pm0.013_{stat}\pm0.006_{sys})\%.
\end{array}\]
As expected in view of the small value of $B_5$,
the branching ratios $B_1$  and $B_3$ are almost fully anti-correlated
with a coefficient of $-0.98$. $B_3$ and $B_5$ have a correlation
coefficient of $-0.08$, and $B_1$ and $B_5$ also have a correlation
coefficient of $-0.08$.   The $B _1$ and $B_3$  results are consistent
with  and slightly  more  precise  than the  results  obtained in  the
PDG~\cite{pdg2000}   combined   fit   to   all  $\tau$   decay   data:
$B_1=(85.32\pm0.13)\%$; $B_3=(14.58\pm0.13)\%$.  The  $B _1$ and $B_3$
results  are  more  than  twice  as  precise  as  the  existing  world
averages~\cite{pdg2000}  of $(84.59\pm0.33)\%$  and $(14.63\pm0.25)\%$
respectively.   The  results are  in  reasonable  agreement with,  but
significantly more precise than,  the most recent direct measurements,
by OPAL~\cite{opaloneprong,opalthreeprong}, CLEO~\cite{cleothreeprong}
and ALEPH~\cite{alephbr1}.

The result  on the 5-prong branching  ratio is in  good agreement with
the world average  of $(0.107\pm0.009)\%$ and the PDG  best fit result
of    $(0.099\pm0.007)\%$ which include contributions    from
OPAL~\cite{opalfiveprong},        CLEO~\cite{cleofiveprong}        and
ALEPH~\cite{alephbr2} measurements.

\subsection*{Acknowledgements}
\vskip 3 mm
 We are greatly indebted to our technical 
collaborators, to the members of the CERN-SL Division for the excellent 
performance of the LEP collider, and to the funding agencies for their
support in building and operating the DELPHI detector.\\
We acknowledge in particular the support of \\
Austrian Federal Ministry of Education, Science and Culture,
GZ 616.364/2-III/2a/98, \\
FNRS--FWO, Flanders Institute to encourage scientific and technological 
research in industry (IWT), Belgium,  \\
FINEP, CNPq, CAPES, FUJB and FAPERJ, Brazil, \\
Czech Ministry of Industry and Trade, GA CR 202/99/1362,\\
Commission of the European Communities (DG XII), \\
Direction des Sciences de la Mati$\grave{\mbox{\rm e}}$re, CEA, France, \\
Bundesministerium f$\ddot{\mbox{\rm u}}$r Bildung, Wissenschaft, Forschung 
und Technologie, Germany,\\
General Secretariat for Research and Technology, Greece, \\
National Science Foundation (NWO) and Foundation for Research on Matter (FOM),
The Netherlands, \\
Norwegian Research Council,  \\
State Committee for Scientific Research, Poland, 2P03B06015, 2P03B11116 and
SPUB/P03/DZ3/99, \\
JNICT--Junta Nacional de Investiga\c{c}\~{a}o Cient\'{\i}fica 
e Tecnol$\acute{\mbox{\rm o}}$gica, Portugal, \\
Vedecka grantova agentura MS SR, Slovakia, Nr. 95/5195/134, \\
Ministry of Science and Technology of the Republic of Slovenia, \\
CICYT, Spain, AEN99-0950 and AEN99-0761,  \\
The Swedish Natural Science Research Council,      \\
Particle Physics and Astronomy Research Council, UK, \\
Department of Energy, USA, DE--FG02--94ER40817.


\begin{thebibliography}{999}
 

\bibitem{pdg96}
See discussion in Particle Data Group, R. Barnett et al., Phys. Rev. {\bf D54} (1996) 1.
\bibitem{cellobr}
CELLO Collab., H. J. Behrend et al., Z. Phys. {\bf C46} (1990) 537.
\bibitem{alephbr1}
ALEPH Collab., D. Decamp et al., Z. Phys. {\bf C54} (1992) 211.
\bibitem{alephbr2}
ALEPH Collab., D. Buskulic et al., Z. Phys. {\bf C70} (1996) 579.
\bibitem{pdg2000}
Particle Data Group, D. E. Groom et al., Eur. Phys. J. {\bf C15} (2000) 1. 
\bibitem{detect}
DELPHI Collab., P. Aarnio et al., Nucl. Instr. and Meth. {\bf A303} (1991) 233.
\bibitem{delphi_performance}
DELPHI Collab., P. Abreu et al.,  Nucl. Instr. and Meth. {\bf A378} (1996) 57.
\bibitem{jadach}
 S. Jadach et al., Comp. Phys. Comm. {\bf 79} (1994) 503;\\
 S. Jadach et al., Comp. Phys. Comm. {\bf 70} (1992) 69.
\bibitem{dymu3}
 J. E. Campagne and R. Zitoun, Z. Phys. {\bf C43} (1989) 469.
\bibitem{babamc}
 F. A. Berends, R. Kleiss and W. Hollik,
 Nucl. Phys. {\bf B304} (1988) 712.
\bibitem{bhwide}
S. Jadach et al., Phys. Lett. {\bf B390} (1997) 298.
\bibitem{sjostrand}
T. Sj\"ostrand, Comp. Phys. Comm. {\bf 82} (1994) 74.
\bibitem{berends}
 F. A. Berends, P. H. Daverveldt and R. Kleiss,
 Phys. Lett. {\bf B148} (1984) 489; \\
 Comp. Phys. Comm. {\bf 40} (1986) 271.
\bibitem{twogam}
T. Alderweireld et al., CERN Report CERN-2000-009 (2000) 219.
\bibitem{tauola}
S. Jadach et al., Comp. Phys. Commun. {\bf 76} (1993) 361.
\bibitem{cleosevenlimit}
CLEO Collab., K. W. Edwards et al., Phys. Rev. {\bf D56} (1997) R5297. 
\bibitem{delphileptonicbr99}
 DELPHI Collab., P. Abreu et al., Eur. Phys. J. {\bf C10} (1999) 201.
\bibitem{lep2xsec}
 DELPHI Collab., P. Abreu et al., Eur. Phys. J. {\bf C11} (1999) 383.
\bibitem{lineshape}
 DELPHI Collab., P. Abreu et al., Eur. Phys. J. {\bf C16} (2000) 371.
\bibitem{btagging}
G. Borisov and C. Mariotti, Nucl. Inst. Meth. {\bf A372} (1996) 181.
\bibitem{delphipizero}
DELPHI Collab., W. Adam et al.,
Z. Phys. {\bf C69} (1996) 561.
\bibitem{bjarnemugam}
DELPHI Collab., P. Abreu et al., Phys. Lett. {\bf B359} (1995) 411.
\bibitem{delphithreepi}
DELPHI Collab., P. Abreu et al., Phys. Lett. {\bf B426} (1998) 411. 
\bibitem{cleostruct}
CLEO Collab., K. W. Edwards et al., Phys. Rev. {\bf D61} (2000) 072003.
\bibitem{delphitaupolarisation}
DELPHI Collab., P. Abreu et al., Eur. Phys. J. {\bf C14} (2000) 585
\bibitem{opaloneprong}                 
OPAL Collab., P. D. Acton et al., Phys. Lett. {\bf B281} (1992) 405. 
\bibitem{opalthreeprong}                 
OPAL Collab., R. Akers et al.,  Z. Phys.  {\bf C66} (1995) 31. 
\bibitem{cleothreeprong}                
CLEO Collab., R. Balest et al., Phys. Rev. Lett. {\bf 75} (1995) 3809.
\bibitem{opalfiveprong}                 
OPAL Collab., P. D. Ackerstaff et al., Eur. Phys. J. {\bf C8} (1999) 183. 
\bibitem{cleofiveprong}                
CLEO Collab., D. Gibaut et al., Phys. Rev. Lett. {\bf 73} (1994) 934.
  
\end{thebibliography}
\end{document}